\documentclass[journal]{IEEEtran}

\ifCLASSINFOpdf
\else
\fi
\usepackage{bm, color}
\usepackage{amsmath}
\usepackage{amssymb}
\usepackage{algorithm}
\usepackage{algpseudocode}
\usepackage{dirtytalk}
\usepackage{multirow}
\usepackage{graphics}
\usepackage{epstopdf}
\usepackage{cite}
\usepackage[dvipsnames]{xcolor}
\usepackage{xr}
\usepackage{cite}

\usepackage{tabulary}
\usepackage{booktabs}
\usepackage{mathtools}
\newtheorem{theorem}{Theorem}[section]

\newtheorem{lemma}[theorem]{Lemma}

\makeatletter
\newcommand*{\addFileDependency}[1]{% argument=file name and extension
  \typeout{(#1)}
  \@addtofilelist{#1}
  \IfFileExists{#1}{}{\typeout{No file #1.}}
}
\makeatother
 
\newcommand*{\myexternaldocument}[1]{%
    \externaldocument{#1}%
    \addFileDependency{#1.tex}%
    \addFileDependency{#1.aux}%
}

\myexternaldocument{supplementary}
\newcommand{\revisions}[1]{{\color{black}{#1}}}

\begin{document}
\title{Covariance Estimation from Compressive Data Partitions using a Projected Gradient-based Algorithm}
%\author{Jonatha Monsalve,
%        %Samuel Pinilla,~\IEEEmembership{Student Member,~IEEE,}
%        XXXX XXXXX,~\IEEEmembership{Member,~IEEE,}
%        XXXX XXXXXXX,~\IEEEmembership{Senior Member,~IEEE}% <-this % stops a space
\author{Jonathan Monsalve,
        Juan Ramirez,
        I\~naki Esnaola,~\IEEEmembership{Member,~IEEE}
        Henry Arguello,~\IEEEmembership{Senior Member,~IEEE \vspace{-8mm}}% <-this % stops a space

\thanks{The work of J. Monsalve was supported by the Colciencias/Department of Santander Scholarship (771 of 2016). J. Monsalve is with the Department of Electrical Engineering, Universidad Industrial de Santander, Bucaramanga 680002, Colombia. J. Ramirez is with the Department of Computer Science, Universidad Rey Juan Carlos, Madrid, 28933, Spain. I. Esnaola is with Department of Automatic Control and Systems Engineering, The University of Sheffield, Western Bank, Sheffield, UK. H. Arguello is with the Department Systems Engineering and Informatics, Universidad Industrial de Santander, Bucaramanga 680002, Colombia(e-mail: henarfu@uis.edu.co)}% <-this % stops a space
%\thanks{Manuscript accepted 01 Feb. 2020; revised xxxx xx, xxxx.}
}

%    \caption{Comparison of the fourth eigenvector of the estimated covariance matrix with: (left) non-filtered gradient and first eigenvector of the bias term. (right) Filtered gradient and the fourth eigenvector of the truth covariance matrix. }

\markboth{IEEE Transactions on Image Processing, XXX~202X}%
{Shell \MakeLowercase{\textit{et al.}}: Bare Demo of IEEEtran.cls for IEEE Journals}
\maketitle

\begin{abstract}
Compressive covariance estimation has arisen as a class of techniques whose aim is to obtain second-order statistics of stochastic processes from compressive measurements. Recently, these methods have been used in various image processing and communications applications, including denoising, spectrum sensing, and compression. Notice that estimating the covariance matrix from compressive samples leads to ill-posed minimizations with severe performance loss at high compression rates. In this regard, a regularization term is typically aggregated to the cost function to consider prior information about a particular property of the covariance matrix. Hence, this paper proposes an algorithm based on the projected gradient method to recover low-rank or Toeplitz approximations of the covariance matrix from compressive measurements. The proposed algorithm divides the compressive measurements into data subsets projected onto different subspaces and accurately estimates the covariance matrix by solving a single optimization problem assuming that each data subset contains an approximation of the signal statistics.
Furthermore, gradient filtering is included at every iteration of the proposed algorithm to minimize the estimation error. The error induced by the proposed splitting approach is analytically derived along with the convergence guarantees of the proposed method. The proposed algorithm estimates the covariance matrix of hyperspectral images from synthetic and real compressive samples. Extensive simulations show that the proposed algorithm can effectively recover the covariance matrix of hyperspectral images from compressive measurements with high compression ratios ($8-15\%$ approx) in noisy scenarios. Moreover, simulations and theoretical results show that the filtering step reduces the recovery error up to twice the number of eigenvectors. Finally, an optical implementation is proposed, and real measurements are used to validate the theoretical findings.
\end{abstract}


\IEEEpeerreviewmaketitle

\section{Introduction}

\IEEEPARstart{C}{ovariance} \revisions{matrix estimation is a statistical problem playing a central role in various signal processing and machine learning applications \cite{Haykin:2002, bishop2006PRML}. For instance, principal component analysis (PCA) is a technique widely used in signal denoising and dimensionality reduction (DR), whose transform is derived from the eigendecomposition of the covariance matrix \cite{Jolliffe86}. However, this estimation task is a data-dependent problem that demands acquisition systems with broad storage capabilities and high computing power. These drawbacks have motivated the proposal of various acquisition schemes that attempt to recover relevant signal information by capturing a reduced number of samples.}

%\revisions{Today, a large number of high-dimensional signals are continuously acquired in the context of different applications such as wireless communication networks \cite{Dai2019Big}, airborne or satellite-borne imaging \cite{agapiou2017remote}, medical imaging \cite{Amirhessam2019Big}, among others. In general, these signals are captured by sensors based on the Nyquist-Shannon sampling theorem \cite{oppenheim2001discrete}, demanding broad storage and processing capabilities of the acquisition systems.} 

\revisions{A representative example is the class of sensors based on the compressive sensing (CS) theory \cite{Donoho2006Compressed, Baranuik2007Compressive, Candes2008Introduction}, whose recovery algorithms assume that the signal of interest admits a sparse representation in a predefined transform domain. Furthermore, it has been proven that these recovery algorithms can obtain the target signal with high probability from CS measurements when the sampling operator is modeled using a random matrix whose entries follow either Gaussian or Bernoulli distributions \cite{rauhut2010compressive}. Seizing the advances in the CS theory, compressive spectral imaging (CSI) has emerged as an acquisition framework that acquires and compresses hyperspectral images simultaneously \cite{CaoComputational2016}. In this context, the coded aperture snapshot spectral imaging (CASSI) optical architecture can be considered the most distinctive CSI sensor \cite{WagadarikarSingle2008}. Different variants of the CASSI architecture with their respective implementations have also been reported, including the dual dispersive CASSI (DD-CASSI) \cite{Gehm2007}, the colored CASSI (C-CASSI) \cite{ArguelloColored2014}, and the spatial-spectral encoded CSI (SSCSI) \cite{Lin2014}. However, the target spectral images do not often exhibit a sparse representation in a computationally tractable transform basis. Additionally, CSI systems typically involve binary sensing matrices that affect the performance of the conventional covariance matrix estimation techniques \cite{8937037}.}

On the other hand, compressive covariance sampling (CCS) has emerged as an acquisition framework to obtain second-order statistics of stochastic processes from compressive measurements \cite{Romero2016Compressive}. This approach has been used in multiple signal processing and communications applications, including spectrum sensing \cite{Ariananda2012Compressive, Qin2017Generalized}, system identification \cite{Testa2016Compressive}, and phase retrieval \cite{Chen2015Exact}. Due to the vast amount of information collected by imaging spectrometers, a few compressive acquisition schemes have been proposed to recover hyperspectral images based on PCA representations. These sensing schemes shift the computational burden from resource-constrained sensors to powerful base stations. For example, compressive projection principal component analysis (CPPCA) estimates both the principal components of the target hyperspectral image and an approximation of the PCA transform matrix from random compressive projections \cite{fowler2009compressive}. In addition, the spectral compressive acquisition (SpeCA) consists of an acquisition scheme and a reconstruction algorithm that recovers the PCA coefficients of hyperspectral vectors from compressive samples \cite{Martin2016}. \revisions{However, most of these methods do not explore a suitable optical implementation that can be used in real scenarios}.

%On the other hand, compressive spectral imaging (CSI) has arisen as an acquisition framework based on the compressive sensing theory \cite{Donoho2006Compressed, Baranuik2007Compressive, Candes2008Introduction} that recovers the spectral image of interest using a reduced number of camera snapshots. This acquisition framework recovers the hyperspectral image by solving a regularized optimization problem whose formulation assumes that the target image is sparsely represented in a given dictionary. In this regard, the coded aperture snapshot spectral imaging (CASSI) optical architecture has been considered the more distinctive CSI system \cite{WagadarikarSingle2008}. Different variants of the CASSI system with their respective practical implementations have been also reported including the dual dispersive CASSI (DD-CASSI) \cite{Gehm2007}, the colored CASSI (C-CASSI) \cite{ArguelloColored2014}, and the spatial-spectral encoded CSI (SSCSI) \cite{Lin2014}. These image acquisition systems typically involve binary sensing matrices that affect the performance of the conventional covariance matrix estimation techniques \cite{8937037}.

%\subsection{Contribution}

This work focuses on developing an algorithm based on the projected gradient method to estimate the covariance matrix from compressive measurements. To this end, compressive measurements are divided into data subsets and projected onto multiple subspaces to improve the condition of the problem. Expressly, the estimation problem aims at recovering a low-rank or Toeplitz representation of a positive semidefinite matrix that minimizes the Frobenius norm of the projection errors. The proposed algorithm is evaluated for estimating the covariance matrix embedded in hyperspectral image signatures using different compressive acquisition schemes, including random projections and binary encoding. It should note that, although the proposed method has been mainly tested on compressive samples derived from hyperspectral images, it can be extended to other image processing and communications applications. \revisions {The contributions of this paper are summarized as follows: i) This paper proposes an optimization problem and a projected gradient-based covariance estimation method from compressive measurements. The proposed method splits compressive samples into partitions projected on different subspaces to improve the estimation accuracy. The lower bound of the optimal number of partitions to obtain a reliable covariance matrix estimation is also derived (Lemma \ref{lemma:singular}). ii) Moreover, this work derives theoretical guarantees for the global convergence of the proposed algorithm and determines the error term induced by the data splitting approach (Lemma \ref{lema:bias}). Likewise, a filtering strategy is proposed to mitigate the error induced by this error. iii) Finally, an implementable sensing protocol based on the DD-CASSI optical architecture is proposed and tested in the lab.

}
\vspace*{-0.3cm}
\subsection{Related work}
Covariance matrix estimation techniques from compressive samples have been reported for hyperspectral images. These methods have been used in different applications such as image reconstruction \cite{fowler2011reconstructions, Li2013Integration}, anomaly detection \cite{Fowler2012Anomaly}, and image classification \cite{Li2013Classification}. For example, the CPPCA approach obtains the PCA coefficients and an approximation of the PCA basis from random compressive projections \cite{fowler2009compressive}.  Notice that CPPCA assumes that the eigenvalues of the target covariance matrix exhibit a highly eccentric distribution. This assumption often does not hold for small eigenvalues in hyperspectral imaging due to the high degree of correlation among the spectral signatures. Secondly, the SpeCA approach introduces a spectral image recovery algorithm tailored to a particular sensor \cite{Martin2016}. More precisely, this method recovers the principal components of images by using a linear mixture model. Notice that the reported sensor requires sensing the entire image before obtaining the random compressive projections. Bioucas et al. proposed COVALSA \cite{Bioucas2014Covalsa}, an algorithm based on the ADMM approach that estimates the covariance matrix from compressive measurements assuming different structures such as Toeplitz, sparseness, and low rank. However, this method approximates the inverse function to use the ADMM approach adding extra hyperparameters. \revisions{Our approach estimates the covariance matrix from compressive samples without resorting to assumptions about the PCA coefficients compared to previous methods. Furthermore, the proposed approach can estimate the covariance matrix for a broader range of sampling operators, including random projections and CSI samples. In contrast to CPPCA and SpeCA, our method is evaluated using real compressive measurements captured by a practical optical setup. In addition, we analytically obtain the optimal number of partitions that recovers a reliable estimation.}

%Additionally,  they use a single sensing matrix for the whole image which is infeasible in real scenarios.\revisions{ Most of these methods are applied to the spectral imaging area, but they fail in providing a practical optical setup to acquire the compressed measurements. In addition, some of them use their own partitions concept, but do not provide a theoretical optimal number of partitions.}

In the context of other applications, various CCS strategies have been reported \cite{Romero2016Compressive}. For example, Romero et al. \cite{Romero2013Compressive} used sparse rulers to recover a Toeplitz version of the covariance matrix. Similar approaches for compressive power spectrum estimation \cite{Ariananda2012Compressive, alwan2021compressive, Qiao2016}, online compressive covariance sampling \cite{park2019online}, and wideband spectrum sensing \cite{Romero2013Sensing} have also been reported. Hanchao Qi and Shannon Hughes\cite{invariance} and Farhad Pourkamali-Anaraki\cite{invariance2} analyzed the bias introduced by Gaussian matrices on the covariance estimation. They proved that the bias depends on the kurtosis of the projection matrix and the dimension of the projected subspace. \revisions{ Azizyan et al.\cite{Azizyan2018} also proposed an unbiased estimator based on properties of the Beta distribution. This method does not exploit the low-rank or Toeplitz structure in covariance matrices, and their accuracy drops for a limited number of samples. Finally, a data-aware covariance estimator from compressive data was recently developed\cite{Chen2020effective,Chen2017}. However, this is not suitable for a compressive sensing set-up since the data must be fully acquired, then a compression matrix is computed by using some properties of the data. Compared to these approaches that build sampling matrices tailored to particular sensing schemes, the proposed approach performs estimations from different operators by splitting the compressive samples into partitions.}
\vspace*{-0.3cm}
\subsection{Paper organization}
The paper is organized as follows: Section \ref{sec:model} introduces the covariance matrix estimation problem from random projections. Section \ref{sec:proposal} presents the optimization problem to be solved and the proposed algorithm for estimating covariance matrices from compressive projections in multiple subspaces. Sections \ref{sec:convergence} and \ref{sec:bias} includes the global convergence guarantee of the proposed algorithm along with the error analysis. In Section \ref{sec:results}, the performance of the proposed algorithm is evaluated using extensive numerical simulations using hyperspectral images. Additionally, an optical implementation is proposed to validate the thoretical findings. Some concluding remarks are summarized in Section \ref{sec:conclusions}.


\section{Compressive covariance sampling formulation}\label{sec:model}

Let $\mathbf{X}=[\mathbf{x}_1,\ldots,\mathbf{x}_n]$ be a matrix whose columns $\mathbf{x}_j \in \mathbb{R}^{l}$ for $j=1,2,\ldots,n$, are independent realizations of a zero-mean Gaussian random vector with covariance matrix  $\bm{\Sigma}$ i.e., the distribution of $\mathbf{x}$ conditioned to $\bm{\Sigma}$ is 
\begin{equation}
    f(\mathbf{x}|\bm{\Sigma}) = \pi^{-l/2} |\bm{\Sigma}|^{-l/2} \text{etr}\left(\frac{1}{2}\bm{\Sigma}^{-1} \mathbf{x}\mathbf{x}^T\right),
\end{equation}
where $\text{etr}(.)$ denotes the exponential of the trace. Under this context, the maximum likelihood estimator (MLE) for the covariance matrix reduces to the sample covariance matrix given by
\begin{equation}
\mathbf{S}=\frac{1}{n}\mathbf{XX}^T=\frac{1}{n}\sum_{j=1}^{n}\mathbf{x}_j\mathbf{x}_j^T,
\label{eq:covfull}
\end{equation}
where $\bm{\Sigma}=\mathbb{E}\left[\mathbf{S}\right] $, $\bm{\Sigma} \in S_{++}^{l \times l}$, with $\mathbb{E}[\cdot]$ denoting the statistical expectation {\color{black} and $S_{++}^{l \times l}$ represents the set of positive definite matrices of size $l\times l$}. However, in many practical applications, lower-dimensional signal projections are available instead of the target high-dimensional signal. In this regard, the sampling process that obtains lower-dimensional signal projections can be modeled as
\begin{equation}
    \mathbf{Y}=\mathbf{P}^T\mathbf{X} + \mathbf{N} = [\mathbf{P}^T\mathbf{x}_1,\cdots, \mathbf{P}^T\mathbf{x}_2] + \mathbf{N},
    \label{eq:randomnoise}
\end{equation}
where $\mathbf{Y} = [\mathbf{y}_1, \mathbf{y}_2, \ldots,\mathbf{y}_n] \in \mathbb{R}^{m\times n}$ is the matrix containing the compressive projections  $\mathbf{y}_j \in \mathbb{R}^{m}$ for $j=1,2,\ldots,n$, $\mathbf{P} \in \mathbb{R}^{l\times m}$ with $m<l$ represents the projection matrix; and $\mathbf{N}\in \mathbb{R}^{m\times n}$ is the additive noise matrix whose entries are characterized as independent and identically distributed (iid) random samples following a zero-mean Gaussian model with variance $\sigma_N^2$, i.e. $N_{i,j}\sim \mathcal{N}(0,\sigma_N^2)$. Notice that the sample covariance matrix obtained from the observation matrix is obtained as
\begin{equation}
    \begin{split}
        \mathbf{\tilde{S}}&=\frac{1}{n}{\mathbf{YY}^T}=\frac{1}{n}(\mathbf{P}^T\mathbf{X} + \mathbf{N})(\mathbf{P}^T\mathbf{X} + \mathbf{N})^T,
        \label{eq:5}
    \end{split}
\end{equation}
with $ \mathbf{\tilde{S}} \in \mathbb{R}^{m\times m}$. Since $\mathbf{x}\sim \mathcal{N}(\mathbf{0},\bm{\Sigma})$ and $\mathbf{P}$ is a fixed matrix, the projected vectors $\{\mathbf{y}_j\}_{j=1}^{n}$ can be modeled as zero-mean Gaussian vectors with covariance given by $\mathbf{P}^T\bm{\Sigma}\mathbf{P} + \sigma_N^2\mathbf{I}$, i.e.  $\mathbf{y} \sim \mathcal{N}(0,\mathbf{P}^T\bm{\Sigma}\mathbf{P} + \sigma_N^2\mathbf{I})$. Furthermore, it can be observed that $n\mathbf{\tilde{S}}$ follows a Wishart distribution, that is, $n\mathbf{\tilde{S}}\sim \mathcal{W}(\mathbf{P}^T\bm{\Sigma}\mathbf{P} + \sigma_N^2\mathbf{I}, n)$\cite{Besson2008}. 

The above assumptions lead to the minimization of the Frobenius norm of the residuals between the covariance matrix of the projected vectors and the projected version of the covariance matrix estimate as the optimal performance criterion. However, this approach leads to ill-posed optimizations with significant performance losses at high compression rates. To overcome this limitation, a regularization term is aggregated to the cost function based on a particular covariance matrix structure, e.g., low-rank or Toeplitz. The optimization problem to recover the sample covariance matrix $\bm{\Sigma}$ from $\mathbf{\tilde{S}}$ is formulated as \cite{Bioucas2014Covalsa} 
\begin{equation}
	\begin{aligned}
	\bm{\Sigma}^* = &\underset{\bm{\Sigma}\in \mathbf{D}}{\text{ argmin}}
	& & \|\mathbf{\tilde{S}}-\mathbf{P}^T\bm{\Sigma}\mathbf{P}\|_F^2 + \tau \psi (\bm{\Sigma}),
	\end{aligned}
	\label{eq:opt1}
\end{equation}
where $\psi(\cdot)$ is a convex function that regularizes the problem, $\tau$ is the regularization parameter, $\|\cdot\|_F$ denotes the Frobenius norm, and $\mathbf{D}$ is a  proper convex and closed set, e.g., the set of positive semi-definitive or Toeplitz matrices. 

Moreover, note that the zero-mean assumption in \eqref{eq:covfull} does not hold in different image processing applications.  Hence, the random projections can be alternatively written as:
\begin{equation}
    \mathbf{Y}=\mathbf{P}^T(\mathbf{X} + \mathbf{\tilde{X}}) + \mathbf{N},
    \label{eq:randomnoise2}
\end{equation}
where $\mathbf{\tilde{X}}=\mathbf{\tilde{x}} \mathbf{1}^T$ is a matrix whose columns are the mean vector, i.e., $\mathbf{\tilde{x}} = \mathbb{E}\left[\mathbf{x}\right]$, and $\mathbf{1} \in \mathbb{R}^{n}$ is an $n$-dimensional vector with one-valued entries. Notice that an estimate of the mean vector can be obtained from the compressive projections \cite{invariance,invariance3} as follows
\begin{equation}
	\mathbf{\tilde{x}}=\alpha \sum_{j=1}^{n} \mathbf{P}(\mathbf{P}^T\mathbf{P})^{-1} \mathbf{y}^j,
	\label{eq:mean}
\end{equation}
where $\alpha=m/n$ and $\mathbf{y}^j$ is the \textit{j-th} vector in  $\mathbf{P}^T\mathbf{X}$. It has been proved in \cite{invariance} that \eqref{eq:mean} converges to the mean vector when $n \rightarrow \infty$. Once the mean is estimated, the measurements can be corrected by subtracting the projection of the estimated mean vector to the biased samples, i.e., $\mathbf{\tilde{Y}}=\mathbf{Y}-\mathbf{P}^T(\mathbf{\widehat{x}} \mathbf{1}^T)$. Without loss of generality, we assume that signals are zero mean.

 \section{Recovery of the covariance matrix from compressed measurements}\label{sec:proposal}
Solving \eqref{eq:opt1} typically yields poor results at high compression ratios of the projection vectors $m/l$. This behavior is attributed to all vectors being projected onto a single subspace and possibly projected onto the null space. Hence, we split the data into disjoint subsets projected onto different subspaces to improve the performance of the estimator. The partitioning into multiple subsets has been previously used for the CPPCA sensing approach\cite{fowler2009compressive}. However, this approach requires that the sensing matrices be orthonormal.
\subsection{Projection set up and optimization problem}
Let's split the dataset $\mathbf{X}$ into $p$ disjoint subsets, $\mathbf{X}_i$, with columns defined as $ \mathbf{X}_i = [\mathbf{x}_{\Omega_{i1}}, \cdots, \mathbf{x}_{\Omega_{ib}}]$ with $i=1,2,\cdots,p$ and $\Omega_{ij}= \Omega_{i'j'}$ only if $i=i'$ and $j=j'$. Since each column $\mathbf{x}_{\Omega_{i1}} \sim \mathcal{N}(0, \bm{\Sigma})$, it holds that the sample covariance matrix of every subset  $b\mathbf{S}_i=\mathbf{X}_i\mathbf{X}_i^T \sim \mathcal{W}(\bm{\Sigma}, b)$, where $b = n/p$ is the number of columns in each subset. Then, each subset $\mathbf{X}_i \in \mathbb{R}^{l\times b}$ is projected in a lower-dimensional subspace with an independent matrix $\mathbf{P}_i \in \mathbb{R}^{l\times m}$, this is,
\begin{equation}
	\mathbf{Y}_i = \mathbf{P}_i^T \mathbf{X}_i + \mathbf{N}_i.
    \label{eq:partition}
\end{equation}
Using this splitting procedure, each $\mathbf{S}_i$ matrix can be estimated solving the optimization \eqref{eq:opt1}, in other words,
\begin{equation}
	\begin{aligned}
	\mathbf{S}_i^* = &\underset{\mathbf{S}_i\in \mathbf{D}}{\text{ argmin}}
	& &  ||\mathbf{\tilde{S}}_i-\mathbf{P}_i^T\mathbf{S}_i\mathbf{P}_i||_F^2 + \tau \psi (\mathbf{S}_i)
	\end{aligned}
	\label{eq:opt2}
\end{equation}
where $\mathbf{\tilde{S}}_i =\frac{1}{b}\mathbf{Y}_i\mathbf{Y}_i^T$ for $i=1,\cdots, p$. Notice that the formulation in \eqref{eq:opt2} involves $p$ different optimization problems, one for each matrix $\mathbf{S}_i$, which increases the number of unknowns and thus the ill-posedness of the problem.  However, note that for a sub-Gaussian process, it holds that
\begin{equation}
    ||\mathbf{S}_i - \bm{\Sigma}||\leq \epsilon,
    \label{eq:assumption1}
\end{equation}
  with probability at least $1 - 2\exp{(-t^2 l)}$, for $b\geq J(t/\epsilon)^2l$, $\epsilon \in (0,1),t \geq 1$, and $J$ depends on the sub-Gaussian norm of $\mathbf{X}$\cite{Vershynin2010}, i.e. the statistics of a subset can approximately describe the statistics of the whole random process.
  
  %i.e. the statistics of a sample (subset) can approximate the statistics of the whole random process. Corollary 5.50 in \cite{Vershynin2010} dictates that if $b=O(l)$, the estimation of the covariance matrix for a sub-Gaussian process is accurate.
 
  \begin{figure}[!htb]
     \centering
     \includegraphics[width=0.4\textwidth]{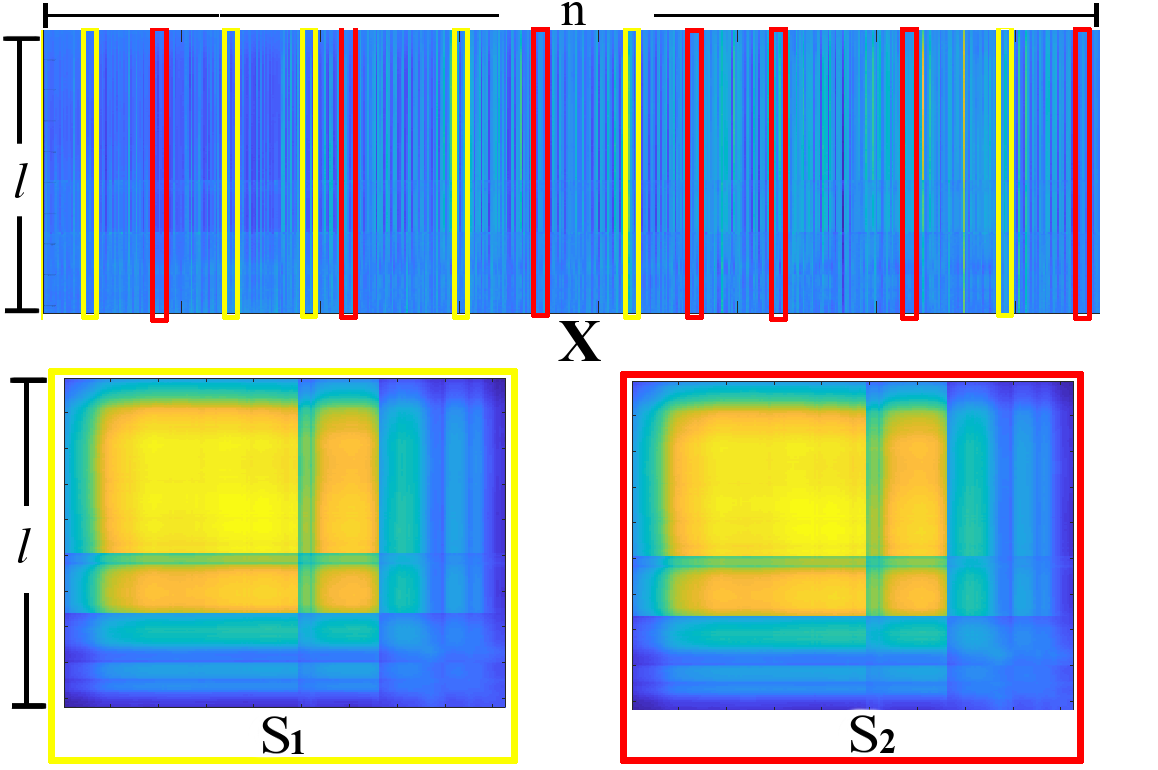}
     \caption{The partition approach. For a matrix $\mathbf{X}$, different subsets of columns are selected, and their covariance matrices are computed. The matrix $\mathbf{S}_1$ represents the covariance matrix of the yellow columns, and $\mathbf{S}_2$ represents the covariance matrix of the red columns.}
     \label{fig:partition}
 \end{figure}
 
 Fig. \ref{fig:partition} illustrates the similarity when the covariance matrices of two subsets of the signal $\mathbf{X}$ are presented. To that end, two subsets of columns of the matrix $\mathbf{X}$ highlighted in yellow and red are used to compute the covariance matrices $\mathbf{S}_1$ and $\mathbf{S}_2$. This example $\mathbf{X}$ is a matrix representation of a hyperspectral image with a spatial resolution of $512\times 512$ (i.e. $n=262144$) and $l=102$ spectral bands, where each column of $\mathbf{X}$ represents the spectrum at a given spatial location. The computation of the matrices $\mathbf{S}_1$ and $\mathbf{S}_2$ uses $b = 2048$ spectral signatures. As it can be seen, these two matrices are similar as $||\mathbf{S}_1-\mathbf{S}_2||_F=0.0321$.
 Instead of recovering all covariance matrices $\mathbf{S}_i$, we assume that $\mathbf{S}_1 = \mathbf{S}_2 = \cdots = \mathbf{S}_p = \bm{\Sigma}$. By doing this, we merge the separate problems in \eqref{eq:opt2} into a single optimization problem by replacing $\mathbf{S}_i$ with $\bm{\Sigma}$, which results in
\begin{equation}
	\begin{aligned}
	\bm{\Sigma}^* = &\underset{\bm{\Sigma} \in \mathbf{D}}{\text{ argmin}}
	& & \sum_{i=1}^{p} ||\mathbf{\tilde{S}}_i-\mathbf{P}_i^T\bm{\Sigma}\mathbf{P}_i||_F^2 + \tau \psi (\bm{\Sigma}).
	\end{aligned}
	\label{eq:opt3}
\end{equation}
Theoretical results and simulations show that this splitting procedure improves the accuracy and variance of the estimator (see Lemma \ref{lemma:singular}).  Additionally, an advantage of \eqref{eq:opt3} in contrast to \eqref{eq:opt1} is that even if an eigenvector falls in the null space of a given matrix $\mathbf{P}_k$, the probability that the eigenvector falls in the null space of every matrix $\mathbf{P}_i$ is small as $m\times p\geq l$, and thus the probability of correct reconstruction increases. To see that, consider a matrix $\mathbf{P} = [\mathbf{P}_1,\mathbf{P}_2,\cdots,\mathbf{P}_p]^T \in \mathbb{R}^{mp\times l}$ whose entries are independent identically distributed subgaussian random variables with zero mean and unit variance. In general, the row null space of the matrix $\mathbf{P}$ is empty if the minimum singular value is greater than 0, i.e., $s_{min}(\mathbf{P})>0$. The probability that the minimum singular value is less than a small number is given by\cite{Vershynin2010}
\begin{equation}
    \begin{aligned}
         \mathbb{P}[\text{null}(\mathbf{P})=\emptyset] &\geq 1 - \mathbb{P}\left(s_{min}(\mathbf{P})\leq \epsilon(\sqrt{mp}-\sqrt{l-1})\right)\\
        &=1-(C\epsilon)^{mp-l+1}+c^{mp}
    \end{aligned}
\end{equation}
with $\epsilon\geq 0$, $C>0$, and $c \in (0,1)$. Note that the probability that the null space to be empty increases exponentially with respect to the subspace dimension $m$ and the number of partitions $p$. Hence, in the case of having a single partition $p$, the only way to increase $\mathbb{P}[\text{null}=\emptyset]$ is to increase the subspace dimension. In the hyperspectral imaging context, increasing $m$ implies acquiring more snapshots reducing the compression. On the other hand, this can also be achieved by increasing the number of partitions $p$ while the compression remains constant.
\vspace*{-0.4cm}
\subsection{Proposed projected gradient algorithm for covariance matrix recovery}
Problem \eqref{eq:opt3} is solved following the projected gradient method. This method requires a differentiable function $f(\bm{\Sigma})$ and a proper closed and convex set $\mathbf{D}$ to formulate the optimization problem as
\begin{equation}
	\begin{aligned}
	\bm{\Sigma}^* = &\underset{\bm{\Sigma}\in \mathbb{R}^{l \times l}}{\text{ argmin}}
	& & f(\bm{\Sigma})\\
    & \text{subject to } 
    & & \bm{\Sigma} \in D.
	\end{aligned}
	\label{eq:opt4}
\end{equation}
Note that \eqref{eq:opt3} has the form of \eqref{eq:opt4}, so it can be solved using the projected gradient algorithm as illustrated in Algorithm \ref{alg:PG}. This algorithm is summarized in three main steps. First, a starting point $\bm{\Sigma}^0 \in D$ and the regularization parameter $\tau$ are selected. Parameter $\tau$ induces the low-rank structure in the solution. Step two (line three), the learning step is selected using the Armijo search\cite{projected}. For that, $\lambda_k= \lambda_{k-1}/\eta^r $, where $\eta > 1$, $\lambda_{-1} > 0$ and $r$ is the smallest positive integer (including 0) that satisfies 
\begin{equation}
    \begin{aligned}
    \underset{\bm{r}}{\text{min }}  f(\bm{\Sigma}^{k_r}, \tau) \leq & f(\bm{\Sigma}^k, \tau) + \text{Tr}(\nabla f(\bm{\Sigma}^k,\tau)^T (\bm{\Sigma}^{k_r})) \\
    + & ||\bm{\Sigma}^{k_r} - \bm{\Sigma}^k||_F^2,
    \end{aligned}
\end{equation}
where $\bm{\Sigma}^{k}$ is the \textit{k-th} iteration, $\bm{\Sigma}^{k_r} =  P_{\mathbf{D}}(\bm{\Sigma}^k - (\lambda_{k-1}/\eta^r) \nabla f(\bm{\Sigma}^k, \tau))$ is an intermediate step between iterations $k$ and $k+1$ and $P_{\mathbf{D}}$ is the projection onto the set $\mathbf{D}$. In step 3 (line 4), the variable $\bm{\Sigma}_k$ is updated by using the gradient of the cost function
\begin{equation}
    f(\bm{\Sigma}, \tau)\equiv \sum_{i=i}^{p} ||\mathbf{\tilde{S}}_i-\mathbf{P}_i^T\bm{\Sigma}\mathbf{P}_i||_F^2 + \tau \psi(\bm{\Sigma}),
    \label{eq:f}
\end{equation}
which for a fixed $\tau$ is given by
\begin{equation}
    \nabla f(\bm{\Sigma}, \tau) = \sum_{i=1}^{p} \mathbf{P}_i(\mathbf{\tilde{S}}_i-\mathbf{P}_i^T\bm{\Sigma}\mathbf{P}_i)\mathbf{P}_i^T + \tau \nabla \psi(\bm{\Sigma}).
    \label{eq:grad1}
\end{equation}

When $\psi(\bm{\Sigma})=\text{Tr}(\bm{\Sigma})$, which is used for low-rank structure, the gradient is given by $\nabla \text{Tr}(\bm{\Sigma})=\mathbf{I} \in \mathbb{R}^{l\times l}$.  Once the variable $\bm{\Sigma}_k$ is updated using the gradient, it is projected onto the set $D$, whose computation depends on the set $D$ itself. This work studies two sets:
\begin{enumerate}
\item  \textbf{Positive semi-definitive:} The orthogonal projection onto the set of positive semi-definitive matrices is given by\cite{Grigoriadis1994}
\begin{equation}
    P_D(\bm{\Sigma}) = \mathbf{W}\bm{\Lambda}_+ \mathbf{W}^T,
    \label{eq:psm}
\end{equation}
where {\color{black}$\mathbf{W}$ is the matrix containing the eigenvectors and} $\bm{\Lambda}_+$ is the matrix containing only the positive eigenvalues of $\bm{\Sigma}$.
\item \textbf{Toeplitz:} The orthogonal projection onto the set of Toeplitz matrices is given by\cite{Grigoriadis1994}
\begin{equation}
    P_D(\bm{\Sigma}) = \begin{bmatrix}
    t_{0}       & t_{1} & t_{2} & \dots & t_{l-1} \\
    t_{1}       & t_{0} & t_{-1} & \dots & t_{l-2} \\
    t_{2}       & t_{1} & t_{0} & \dots & t_{l-3} \\
    \dots & \dots & \dots & \dots & \dots \\
    t_{l-1}       & t_{l-2} & t_{l-3} & \dots & t_{0}
\end{bmatrix},
\label{eq:toepproj}
\end{equation}
where $t_k = 1/(n-k) \sum_{i=1}^{n-k} \Sigma_{i,(i+k)}$, with $\bm{\Sigma} = \{\Sigma_{i,j}\}$.
%\item \textbf{Positive semi-definitive and Toeplitz:} When the set $\mathbf{D}=\mathbf{D}_1 \cap \mathbf{D}_2$, with $\mathbf{D}_1$ the positive semi-definitive set and $\mathbf{D}_2$ the Toeplitz set, the orthogonal projection is found by using Algorithm \ref{alg:PG2} (\cite{Grigoriadis1994,Gubin1967})
%\begin{algorithm}[H]
%\caption{Alternating convex projections}\label{alg:PG2}
%\begin{algorithmic}[1]
%    \State $\bm{\Sigma} \gets \bm{\Sigma}^k - \lambda_k \nabla f(\bm{\Sigma}^k)$ \Comment{\textit{from step 4 in Alg. \ref{alg:PG}}} 
%	\While{stopping criteria is not satisfied}
%	    \State $\bm{\Sigma}_1 \gets P_{D1}(\bm{\Sigma}),\bm{\Sigma}_2 \gets P_{D2}(\bm{\Sigma}_1),\bm{\Sigma}_3 \gets P_{D1}(\bm{\Sigma}_2)$ 
%	    \State $\tau \gets ||\bm{\Sigma}_1 - \bm{\Sigma}_2||^2/ \text{trace}((\bm{\Sigma}_1 - \bm{\Sigma}_3)^T(\bm{\Sigma}_1 - \bm{\Sigma}_2))$
%	    \State $\bm{\Sigma} \gets \bm{\Sigma}_1 + \tau (\bm{\Sigma}_3 -\bm{\Sigma}_1)$
%	\EndWhile
%\end{algorithmic}
%\end{algorithm}

\end{enumerate}

 Thus, the proposed gradient algorithm can be summarized by Alg. \ref{alg:PG}.
\vspace*{-0.3cm}
\begin{algorithm}[H]
\caption{Projected gradient algorithm}\label{alg:PG}
\begin{algorithmic}[1]
	\State $\bm{\Sigma}^0 \in \mathbf{D}, \tau, \lambda_0 $
	\While{stopping criteria is not satisfied}
	    \State $\text{pick } \lambda_k > 0$ \Comment{Armijo search}
	    \State $\bm{\Sigma}^{k+1} \gets P_{\mathbf{D}}(\bm{\Sigma}^k - \lambda_k \nabla f(\bm{\Sigma}^k, \tau))$ \Comment{Using 1) or 2)}
	\EndWhile
\end{algorithmic}
\end{algorithm}
\vspace*{-0.3cm}
Note that, Algorithm \ref{alg:PG} works for both low-rank and Toeplitz cases. Nevertheless, for the Toeplitz case, $\tau$ is set to zero since the low-rank constraint is unnecessary. The algorithm convergence analysis is presented in the Supplementary material, Section \ref{sec:convergence}
 
\section{Error term of the proposed estimator}\label{sec:bias}
The assumption in \eqref{eq:assumption1} introduces an error term in the gradient. To show that, let us characterize the difference of the ground-truth covariance matrix $\mathbf{S}$ and the covariance matrices $\mathbf{S}_i$ as
\begin{equation}
    \mathbf{S}_i = \mathbf{S} + \mathbf{R}_i,
    \label{eq:error}
\end{equation}
where $\mathbf{R}_i \in \mathbb{R}^{l\times l}$ is a matrix that accounts for the error between the covariance matrices. In the ideal case, where $\mathbf{S}=\mathbf{S}_i \forall i$, the estimator is optimal and \eqref{eq:grad1} holds. However, in the more realistic scenario where $\mathbf{S} \neq \mathbf{S}_i \forall i$, assuming \eqref{eq:error}, the error is described in lemma \ref{lema:bias}
\begin{lemma}
    The gradient step for the proposed Algorithm \ref{alg:PG} has an error term given by $\text{Error}[\nabla \tilde{f}(\bm{\Sigma})]=-\sum_{i=1}^p \mathbf{P}_i\mathbf{P}_i^T\mathbf{R}_i\mathbf{P}_i\mathbf{P}_i^T$.
    \label{lema:bias}
    
    where $\text{Error}[\cdot]= \nabla f - \nabla \tilde{f}$, and $\nabla f, \nabla \tilde{f}$ are the optimal and actual gradients respectively.
\end{lemma}
\textit{Proof:} See Appendix \ref{sec:errorproof}.

An important property of the error term is that it is proportional to the number of subsets, and thus the error associated with $\text{Error}[\nabla \tilde{f}(\bm{\Sigma})]$ increases with the number of partitions $p$. However, more partitions improve the condition of the information matrix of the problem. Consequently, choosing the number of subsets is a trade off between improving the condition of the problem and increasing the error. The following theorem bounds the latter.
\begin{theorem}
    The variance for any Covariance matrix $\bm{\Sigma}$ estimator for \eqref{eq:partition} with deterministic projection matrices $\mathbf{P}_i$, assuming that is non-singular, satisfies
    \begin{equation}
\begin{aligned}
     \text{var}(\bm{\tilde{\Sigma}})\geq \frac{p}{n}\text{Tr}\left[\left(\sum_{i=1}^p \mathbf{P}_i\mathbf{A}_i^T\mathbf{P}_i^T\otimes \mathbf{P}_i\mathbf{A}_i\mathbf{P}_i^T\right)^{-1}\right],
\end{aligned}
    \label{eq:cr_theo}
\end{equation}
with $\mathbf{A}_i = (\bm{\Sigma}_N + \mathbf{P}_i^T\bm{\Sigma}\mathbf{P}_i)^{-1}$, and $\left(\sum_{i=1}^p \mathbf{P}_i\mathbf{A}_i^T\mathbf{P}_i^T\otimes \mathbf{P}_i\mathbf{A}_i\mathbf{P}_i^T\right)$ is the information matrix.  \textit{Proof:} See appendix \ref{sec:cramer}.
\label{theo:cramer}
\end{theorem}
From Theorem \ref{theo:cramer} it is important notice that for small values of $p$ the fisher information matrix is singular. Hence, a large enough number of partitions ($p>l^2/m^2$) must be performed based on lemma \ref{lemma:singular}
\begin{lemma}
    Let $\mathbf{P}_i \in \mathbb{R}^{l\times m}$ and $\mathbf{A}_i \in \mathbb{R}^{m\times m}$, then the matrix $\sum_{i=1}^p \mathbf{P}_i\mathbf{A}_i^T\mathbf{P}_i^T\otimes \mathbf{P}_i\mathbf{A}_i\mathbf{P}_i^T$ is singular if $p<l^2/m^2$. \textit{Proof:} See appendix \ref{sec:singular}
    \label{lemma:singular}
\end{lemma}
From Lemma \ref{lemma:singular}, it can be seen that the information matrix is non-singular for some $d > 1$ such that $p \geq d l^2/m^2$. Nevertheless, choosing large $p$ increase the norm of the error term given in Lemma \ref{lema:bias}, {\color{black}as shown in \eqref{eq:bound}}. Hence, $p$ should be chosen big enough to avoid the singularity of \eqref{eq:cr_theo} but small enough to decrease the error term in \eqref{eq:bias}, {\color{black} which yields an optimal number of partitions of $p=\lceil (l^2/m^2)+1 \rceil$}.
Additionally, this error term follows an important property given by lemma \ref{lemma:zero}
\begin{lemma}
    Let $\{\mathbf{R}_i\}$  be the set of error matrices for the subsets covariance matrices $\mathbf{S}_i$, hence since the sensing matrices $\mathbf{P}_i$ are deterministic and $\mathbb{E}[\mathbf{R}]=\mathbf{0}$ (Appendix \ref{sec:proofR}), for any entry $\mathbf{B}_{ij}$ of the matrix $\mathbf{P}_i\mathbf{P}_i^T\mathbf{R}_i\mathbf{P}_i\mathbf{P}_i^T=\mathbf{B}$ it holds that
    \begin{equation}
        \mathbb{E}[\mathbf{B}_{ij}] = 0,
    \end{equation}
    \label{lemma:zero}
    \textit{Proof:} See Appendix \ref{sec:proofentries}.
\end{lemma}

This result motivates the use of a filtered gradient to remove the effect of the error term. Simulations show that this error is usually associated with high frequencies. Moreover, the proposed algorithm filters the gradient in each iteration to mitigate this error, especially when the compression is high since more partitions are required (as can be seen in Lemma \ref{lemma:singular}). The filtered gradient is given by
\begin{equation}
    \nabla \hat{f}(\bm{\Sigma}) = \mathbf{K} * \nabla f(\bm{\Sigma}),
    \label{eq:grad}
\end{equation}
where $*$ represents the convolution operation, and $\mathbf{K} \in \mathbb{R}^{k\times k}$ is the filter kernel. This new gradient is used in step 4 of algorithm \ref{alg:PG}. {\color{black} This filtering step reduces the error term variance, as shown in Appendix \ref{sec:varred} in the supplementary material.  Additionally, the norm of the error term is bounded by:
\begin{equation}
	        \mathbb{P}\left\{ \left\| \sum_{i=1}^{p} \mathbf{H}_i\mathbf{R}_i \mathbf{H}_i\right\|_2 \geq t \right\} \leq 2\times l\times  e^{\frac{-t^2/2}{\sigma_H^2 + \sigma_m^2 \epsilon t/3}}.
	    \end{equation}
\textit{Proof:} See Appendix \ref{sec:bounderror} in supplementary material.
	    }

\section{Simulations and Results}\label{sec:results}
The performance of the proposed algorithm is tested using synthetic and real data. The gradient is filtered using a Gaussian filter with $\sigma = 1$; however, it is only used along with the low-rank restriction (i.e., $\tau >0$).  Three different projection matrices $\mathbf{P}$ are used: \textit{i)} Gaussian matrices whose entries follow a standard normal distribution $\mathbf{P}_{i,j} \sim \mathcal{N}(0,1)$; \textit{ii)} Binary matrices with entries $\mathbf{P}_{i,j} \sim \text{Bernoulli}(p=\frac{1}{3})$; and \textit{iii)}  matrices whose elements obey to a standard uniform distribution, $\{\mathbf{P}\}_{i,j} \sim U(0,1)$. In simulations, two noisy scenarios of 20 and 30 dB SNR were tested with SNR defined as SNR=$10\log{||\mathbf{P}^T\mathbf{X}||_F^2/||\mathbf{N}||_F^2}$. 

\subsection{Synthetic data performance evaluation}
 Synthetic data from a low-rank and Toeplitz covariance matrices were generated. For the low-rank covariance matrix, the data points were generated using Matlab with $\bm{\mu} = 0$, the rank of $\bm{\Sigma}$ set to 7, and the dimension of the signal $l=100$. The data from the Toeplitz matrix was generated as an autoregressive model of order $q=8$ and dimension of the signal $l=100$. For the reconstruction algorithm $\tau = \rho*\text{trace}(\mathbf{S}_0)$, where $\mathbf{S}_0$ is the initialization of the covariance matrix, and $\rho$ was chosen using cross-validation. More details are available in the supplementary material.

\begin{figure}
    \centering
    \includegraphics[width=0.23\textwidth]{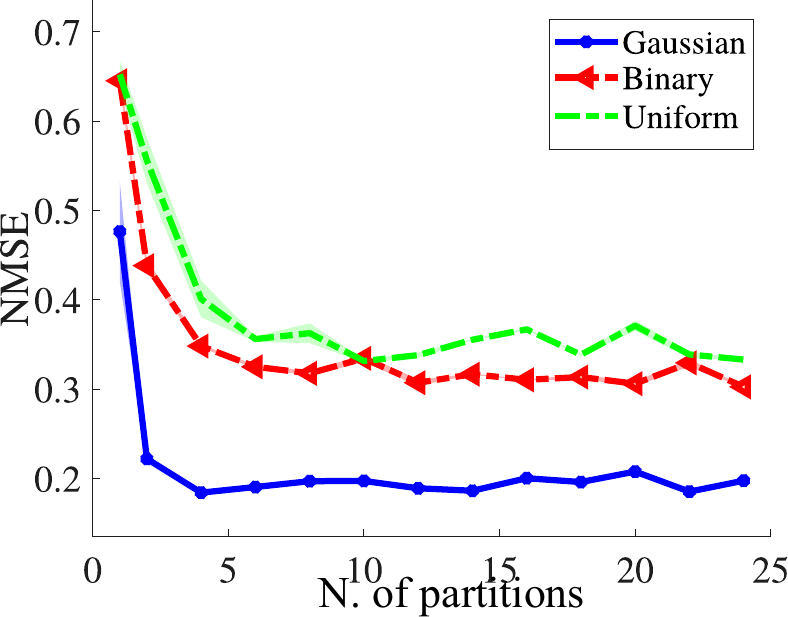}
    \includegraphics[width=0.23\textwidth]{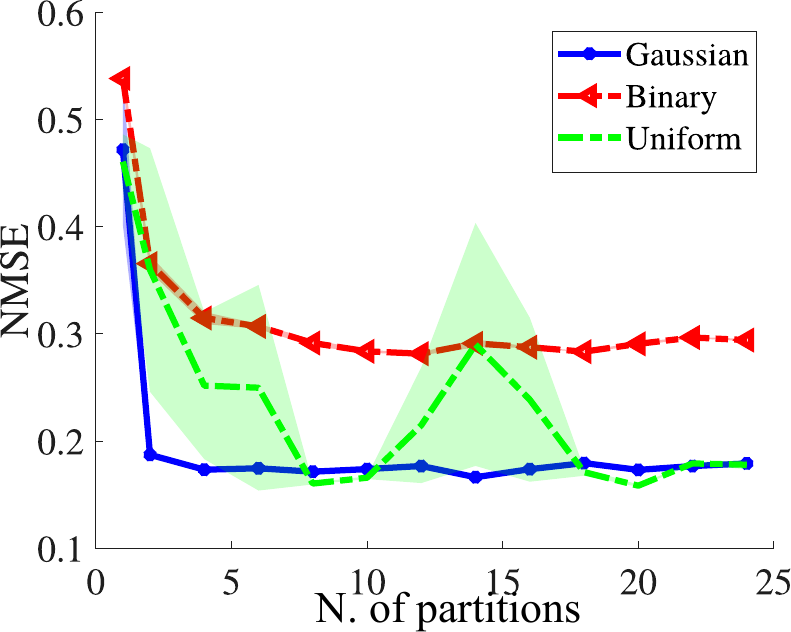}\\
    \includegraphics[width=0.23\textwidth]{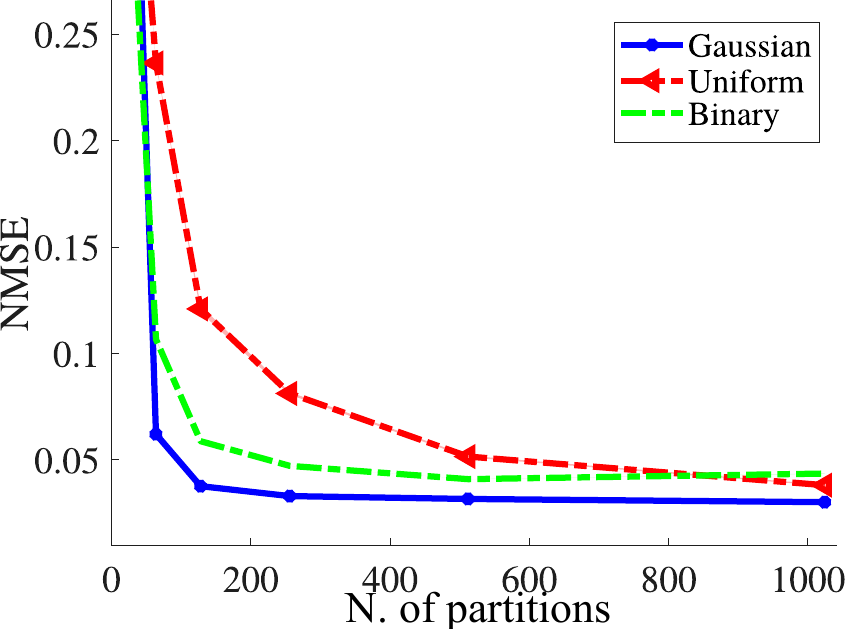}
    \includegraphics[width=0.23\textwidth]{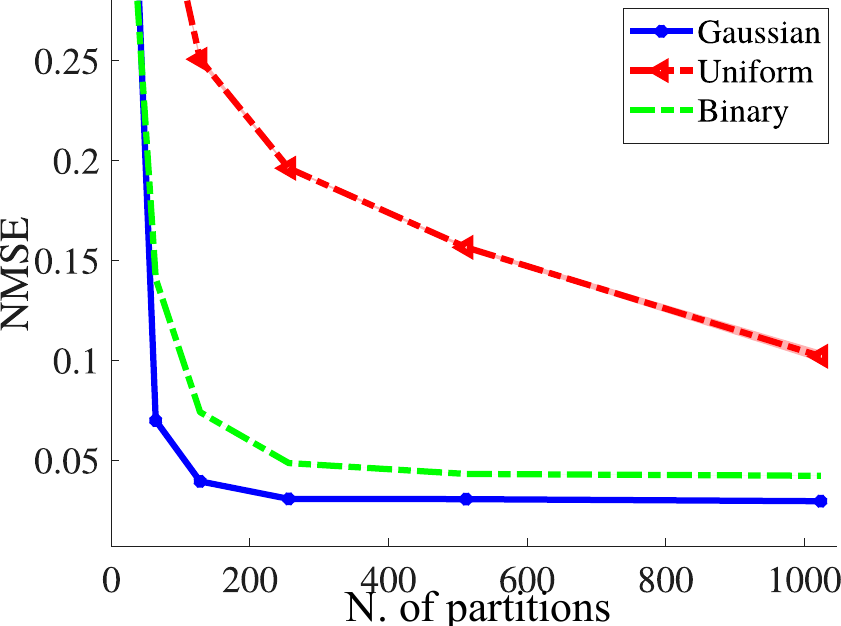}
    \caption{Average normalized mean square error of the reconstructed covariance matrix for the (top) Toeplitz and (bottom) low-rank matrices varying the number of partitions using $8\%$ of compression ratio with two noise scenarios (left) SNR=30dB and (right) SNR=20dB. Each line represents a different sensing matrix (Gaussian, Uniform, Binary). The shaded areas represent the confidence interval (in some cases, it can not be seen in the plot).}
    \label{fig:syntheticpartition}
\end{figure}

Fig. \ref{fig:syntheticpartition} shows the average normalized mean squared error defined as NMSE=$||\bm{\Sigma}-\tilde{\bm{\Sigma}}||_F/||\bm{\Sigma}||_F$ as a function of the number of partitions between the original and reconstructed covariance matrices. It can be seen that Gaussian matrices have the best performance. 
Based on those results, we set the number of partitions to 4 and 128 for Toeplitz and low-rank data, respectively in synthetic data experiments. The proposed algorithm results are compared against sparse rulers and a least squares autoregressive estimator for the Toeplitz matrix. The proposed algorithm is compared against the compressive-projection principal component analysis (CPPCA)\cite{fowler2009compressive} and the spectral compressive acquisition (SpeCA) method for the low-rank matrix\cite{Martin2016}.

\begin{figure}[!htb]
    \centering
    \includegraphics[width=0.23\textwidth]{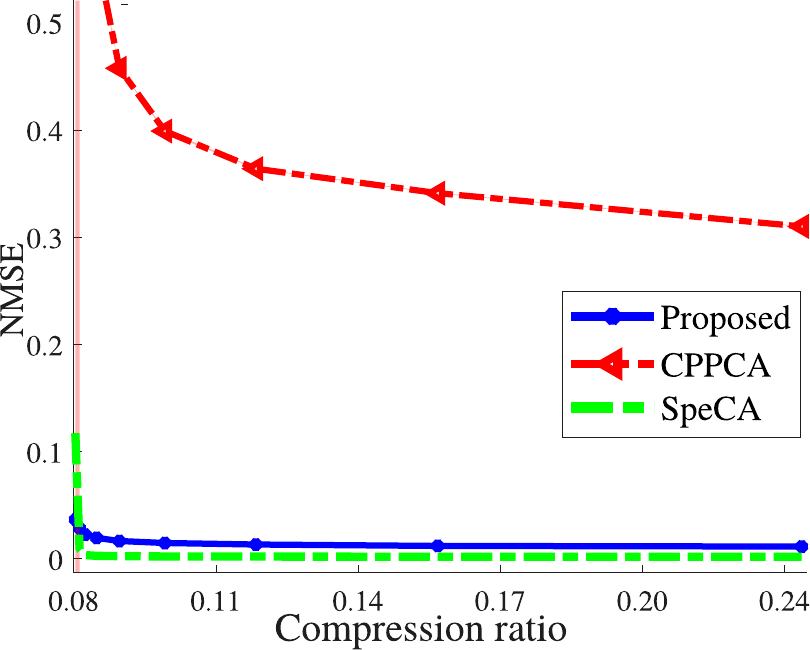}
    \includegraphics[width=0.23\textwidth]{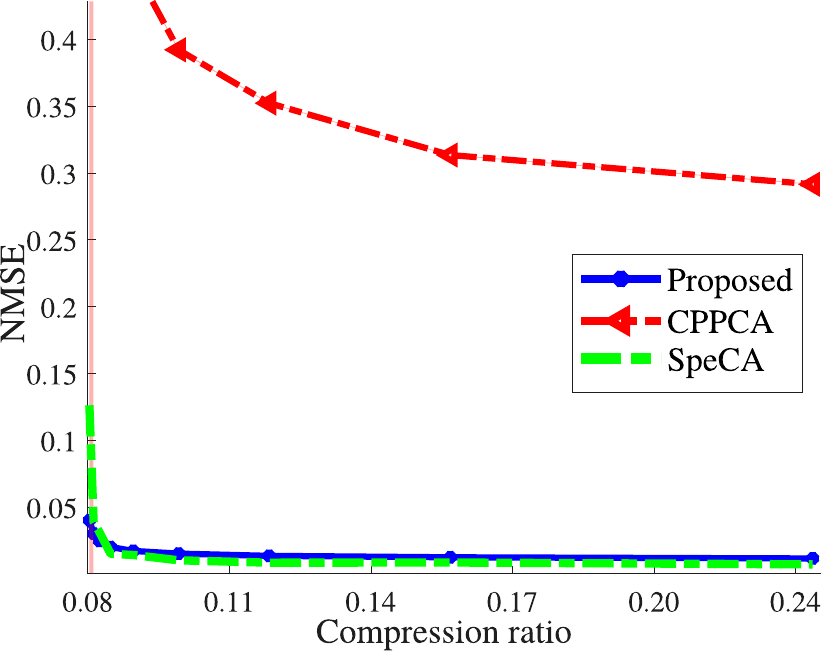}\\
    \includegraphics[width=0.23\textwidth]{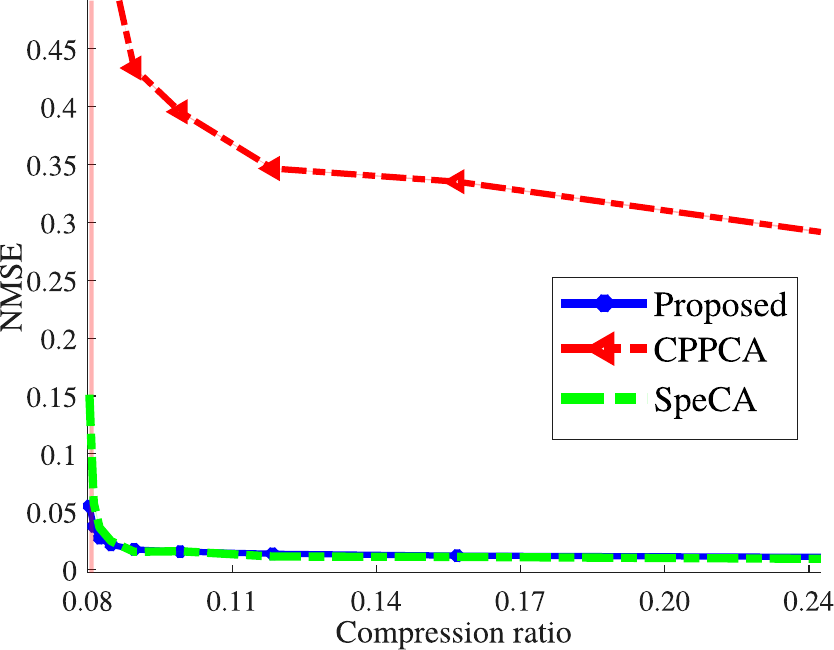}
    \includegraphics[width=0.23\textwidth]{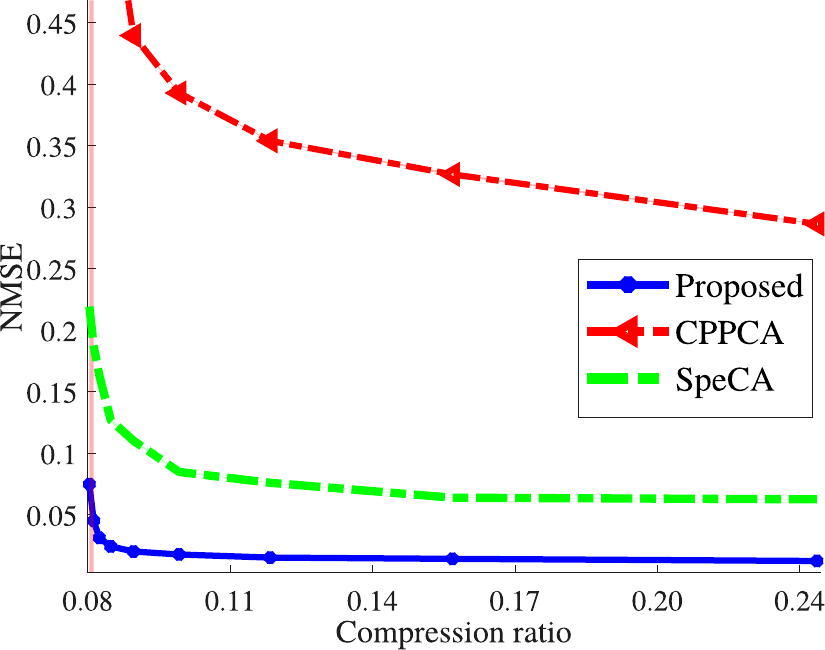}
    \caption{Average normalized mean square error of the reconstructed covariance matrix for the low-rank covariance matrix varying the number of acquisitions using $128$ partitions with two noise scenarios (left) SNR=30dB and (right) SNR=20dB and two types of sensing matrices, Gaussian (top) and Binary(Bottom)}
    \label{fig:syntheticcompression}
\end{figure}
Figure \ref{fig:syntheticcompression} shows that both the proposed and SpeCA algorithms outperform the CPPCA algorithm because the generated random signal does not exhibit an eccentric behavior in the eigenvalues of the covariance matrix which is an essential assumption for the CPPCA algorithm. On  the other hand, the proposed algorithm achieves comparable results to the SpeCA when Gaussian matrices are used but outperforms the SpeCA with binary matrices and in low SNR regimes.

\begin{figure}[!htb]
    \centering
    \includegraphics[width=0.23\textwidth]{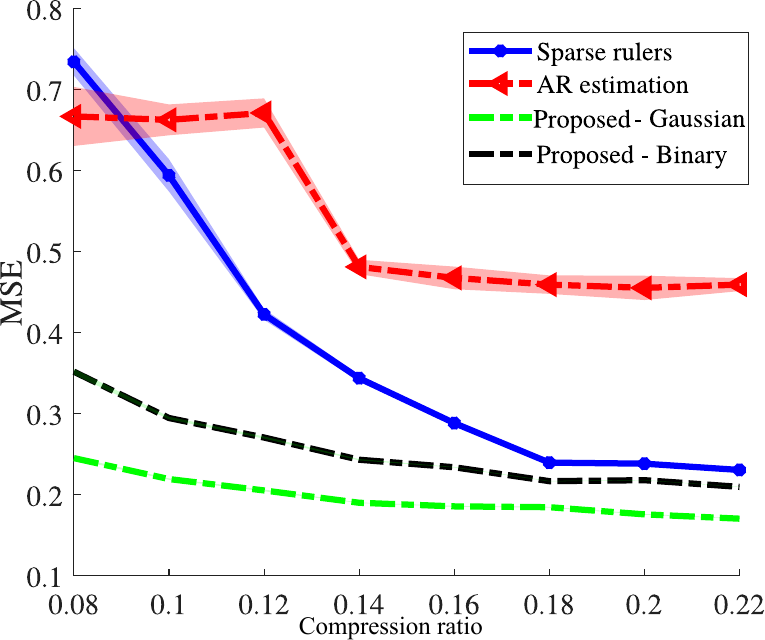}
    \includegraphics[width=0.23\textwidth]{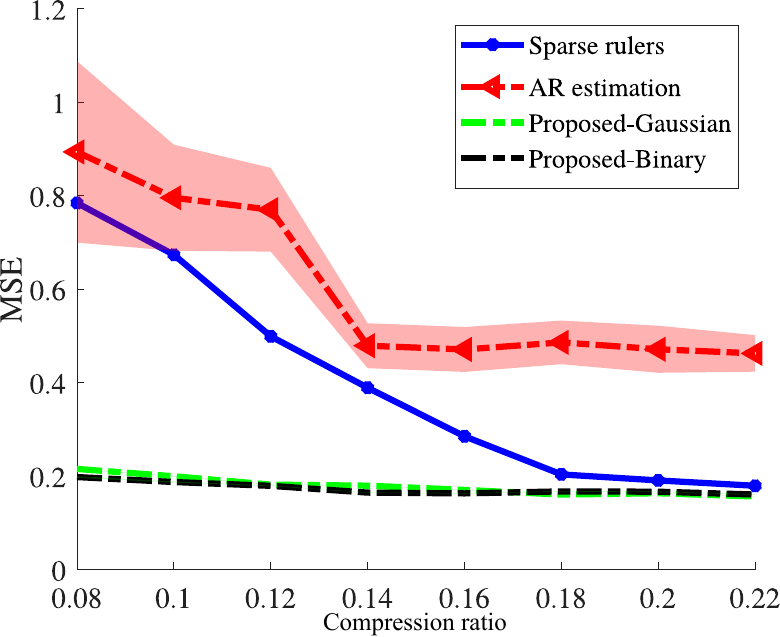}
    \caption{Average normalized mean square error of the reconstructed covariance matrix for the Toeplitz covariance matrix varying the number of acquisitions using four partitions with two noise scenarios (left) SNR=30dB and (right) SNR=20dB. The shaded areas represent the confidence interval.}
    \label{fig:syntheticcompression1}
\end{figure}

Figure \ref{fig:syntheticcompression1} compares the performance of different algorithms in the recovery of the Toeplitz covariance matrix. The proposed algorithm outperforms two state-of-the-art algorithms, AR coefficient\cite{Testa2016Compressive}, and Sparse rulers\cite{Romero2016Compressive}, especially with high compression ratios. The proposed method is compared using two sensing matrices, Gaussian and Binary. Note that both AR coefficients and sparse rulers propose a specific sensing protocol, and hence the sensing matrix is fixed.
\vspace*{-0.1cm}
\subsection{Computational simulations with Hyperspectral images}

\begin{figure}[!htb]
	\centering
	\begin{tabular}{c}
	    \includegraphics[width=0.9\linewidth]{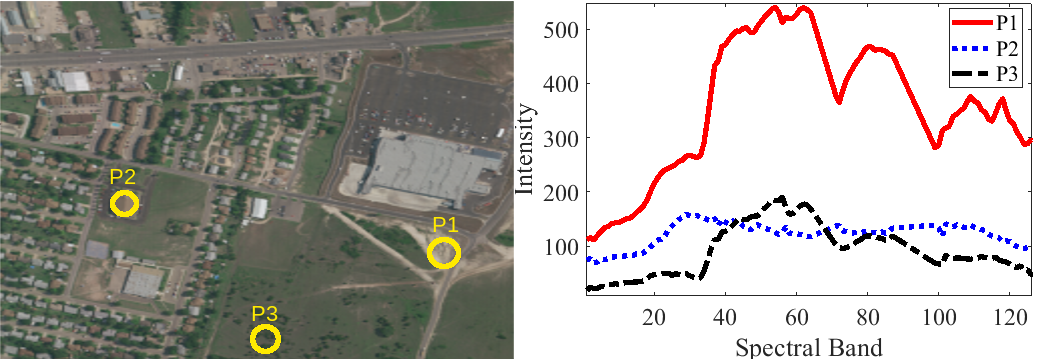}
	    \\
	    \includegraphics[width=0.9\linewidth]{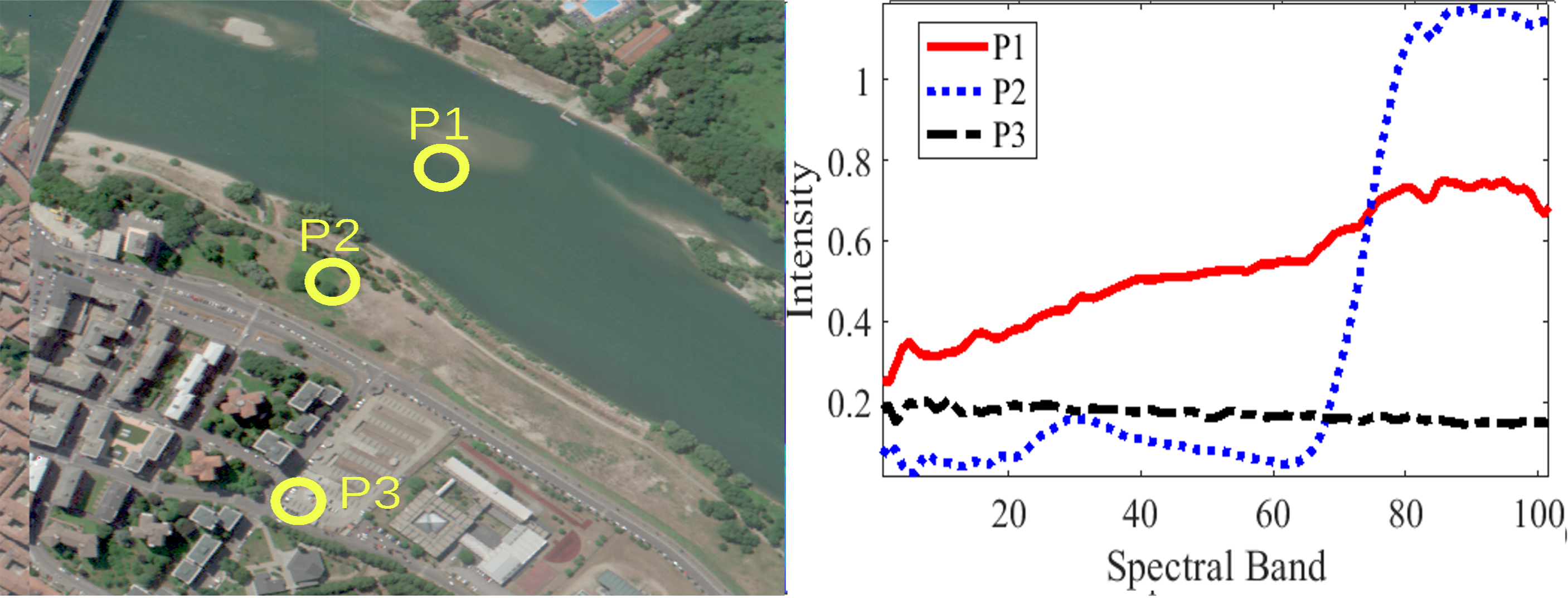}
%	    &
%	    \includegraphics[width=1\linewidth]{Img/pavia_espa.png}
%    \vspace{-.3cm}
    \end{tabular}
	\caption{(Top) Urban dataset: (Left) RGB composite of the hyperspectral image, (Right) spectral signatures of three different pixels at locations \textsf{P1}, \textsf{P2}, and \textsf{P3}.(Bottom) Pavia Centre dataset: (Left) RGB composite of the spectral image, (Right) spectral signatures of three different pixels.}\label{fig:dataset1}
\end{figure}

Additionally, the proposed method is evaluated by estimating the covariance matrix of hyperspectral images using subsets of random compressive projections. Two hyperspectral images are considered: the Urban dataset \cite{dataset} with a spatial resolution of $256\times 256$ pixels and $l=128$ spectral bands; and a section of the Pavia Centre dataset \cite{Dataset2} with dimensions  $512\times 512 \times 102$. The RGB composite and the spectral signatures of three pixels (at the spatial locations \textsf{P1}, \textsf{P2}, and \textsf{P3}) for the Urban dataset are displayed in Fig. \ref{fig:dataset1} (Top-Left) and (Top-Right), respectively. Moreover,  Figs. \ref{fig:dataset1}(Bottom-Left)-(Bottom-Right) show the RGB composite and the spectral signatures for the Pavia Centre dataset. The results obtained with the proposed method are compared with those obtained using the CPPCA and the SpeCA algorithms. Three metrics are used to compare the results, the Mean Square Error (MSE) between the covariance matrices, the error angle between the eigenvectors, and the Peak Signal to Noise Ratio (PSNR). The sample covariance matrix $\mathbf{S}=\mathbf{XX}^T/n$ is used as the truth covariance matrix for the simulations.
\vspace*{-0.2cm}
\subsection{Cramer-rao lower bound and optimal number of partitions}
\begin{figure}[H]
    \centering
    \includegraphics[scale=0.33]{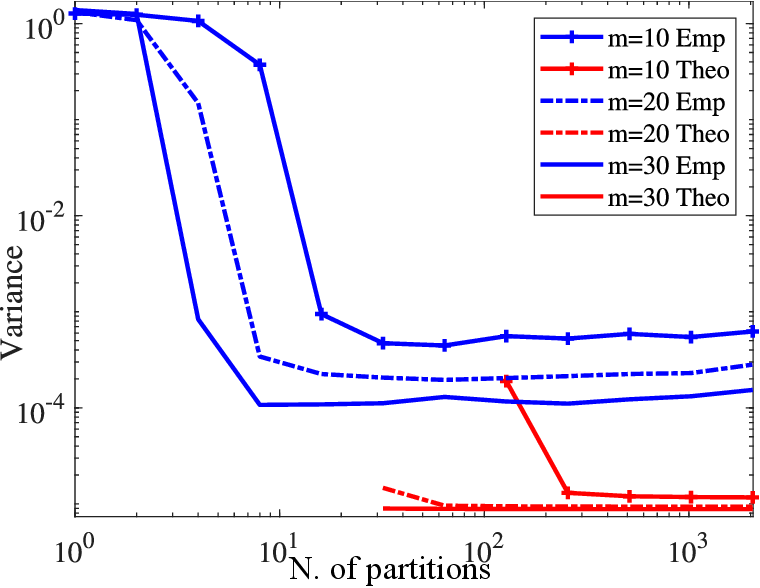}
    \includegraphics[scale=0.33]{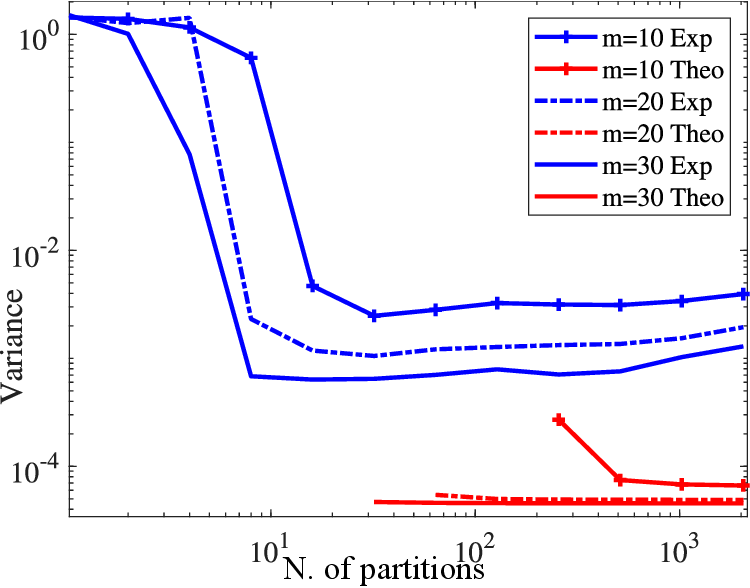}
    \caption{Comparison of the Cramer-rao lower bound and the empirical variance. Blue lines represent the empirical value, and red lines represent the theoretical value. Note that the red lines are shown only when the fisher information matrix is non-singular.}
    \label{fig:cramer_fig}
\end{figure}

As described in Section \ref{sec:proposal}, the signal splits into $p$ subsets projected using different matrices and $p\geq l^2/m^2$ as described by Lemma \ref{lemma:singular}. This section evaluates the estimator's variance using the theoretical expression given in Theorem \ref{theo:cramer} and the empirical variance in the simulations. Table \ref{table:dr} shows the value $l^2/m^2$ for both images as $m$ increases.
\begin{table}[H]
\centering
 \caption{Minimum optimal number of partitions.}
 \label{table:dr}
 \begin{tabular}{|c|c|c|c|c|c|c|c|} 
 \hline
  image/m & 8 & 12 & 16 & 20 & 24 & 28 & 32 \\ %[0.5ex] 
 \hline
 \textbf{Urban }$p$ & 256 & 114 & 64 & 41 & 29 & 21 & 16 \\ 
  \hline
 \textbf{Pavia }$p$  & 163 & 72 & 41 & 26 & 18 & 13 & 10 \\
 \hline
\end{tabular}
\end{table}
Fig. \ref{fig:cramer_fig} shows the theoretical variance given by the Cramer-rao lower bound and the empirical variance defined as $1/r \text{tr}[\sum_{i=1}^r (\bm{\tilde{\sigma}}-\bm{\sigma}) (\bm{\tilde{\sigma}}-\bm{\sigma})^T]$ where $r$ is the number of realizations.
 Fig. \ref{fig:cramer_fig} presents three different compression ratio scenarios going from 6\% to 30\%. It can be seen that the values of $p$ presented in table \ref{table:dr} match those obtained in Fig. \ref{fig:cramer_fig}. Note that the red lines are shown only when the Fisher information matrix is non-singular, which, as expected, is close to the point of most minor empirical variance.
 \vspace*{-0.3cm}
\subsection{Accuracy of the recovered covariance matrix}
\begin{figure}[!htb]
    \centering
    \includegraphics[scale=0.30]{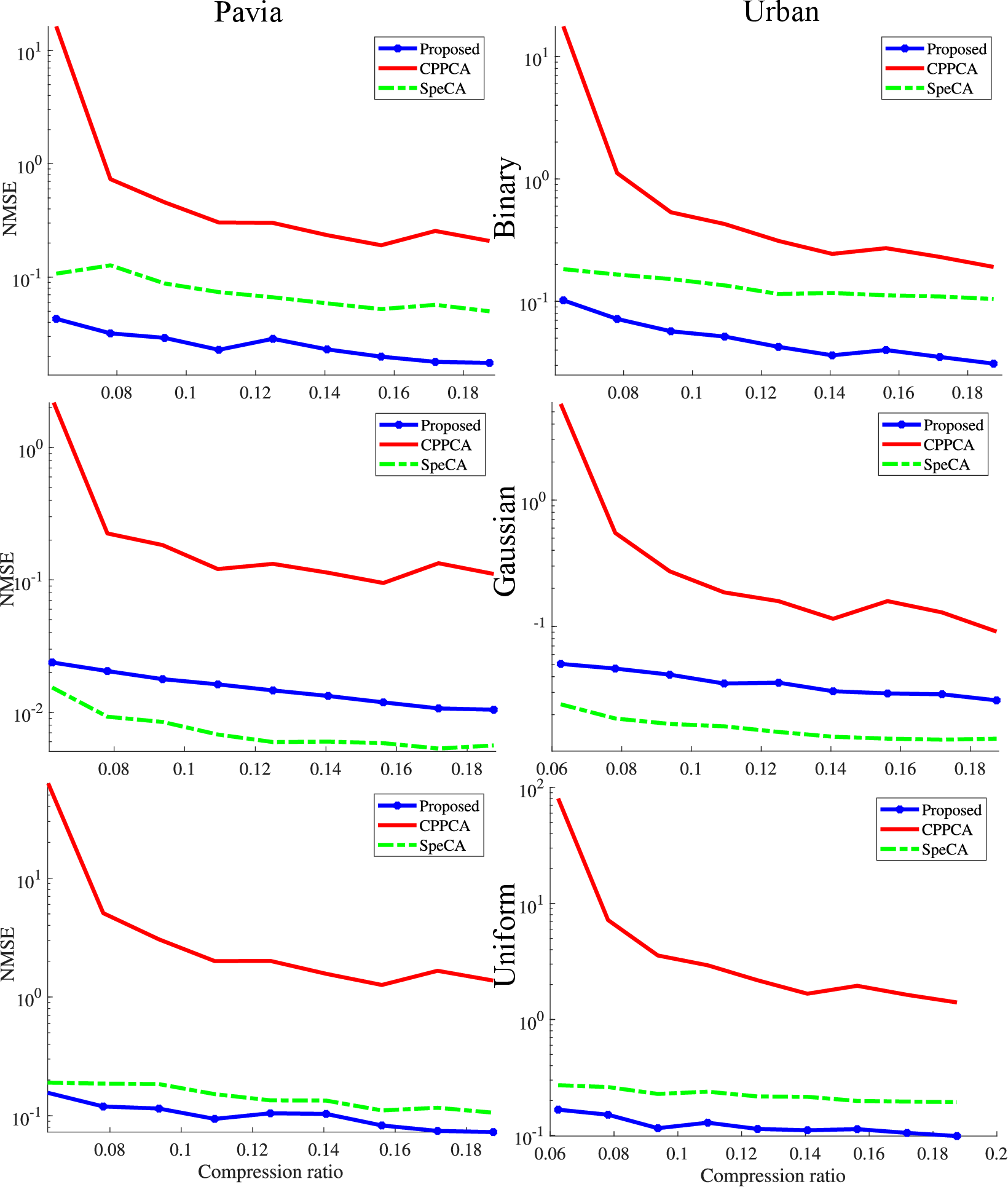}
    \caption{Average MSE of recovered covariance matrix when the compression ratio varies for the Urban image using the number of partitions given in table \ref{table:dr}}
    \label{fig:eigenvectors3}
\end{figure}
The quality of the reconstructed covariance matrices was evaluated using the NMSE and the angle between the eigenvectors of the ground-truth covariance matrix and the recovered eigenvectors using Algorithm \ref{alg:PG}. In Fig. \ref{fig:eigenvectors3}, the NMSE of the reconstructed covariance matrix using different types of matrices is shown. It can be seen that the proposed method outperforms both traditional methods (CPPCA, SpeCA), mainly when binary matrices are used. For the case of Gaussian matrices, $\mathbf{P}$ the proposed algorithm obtain comparable results to SpeCA.

\begin{figure}[!htb]
    \centering
    \includegraphics[width=0.48\textwidth]{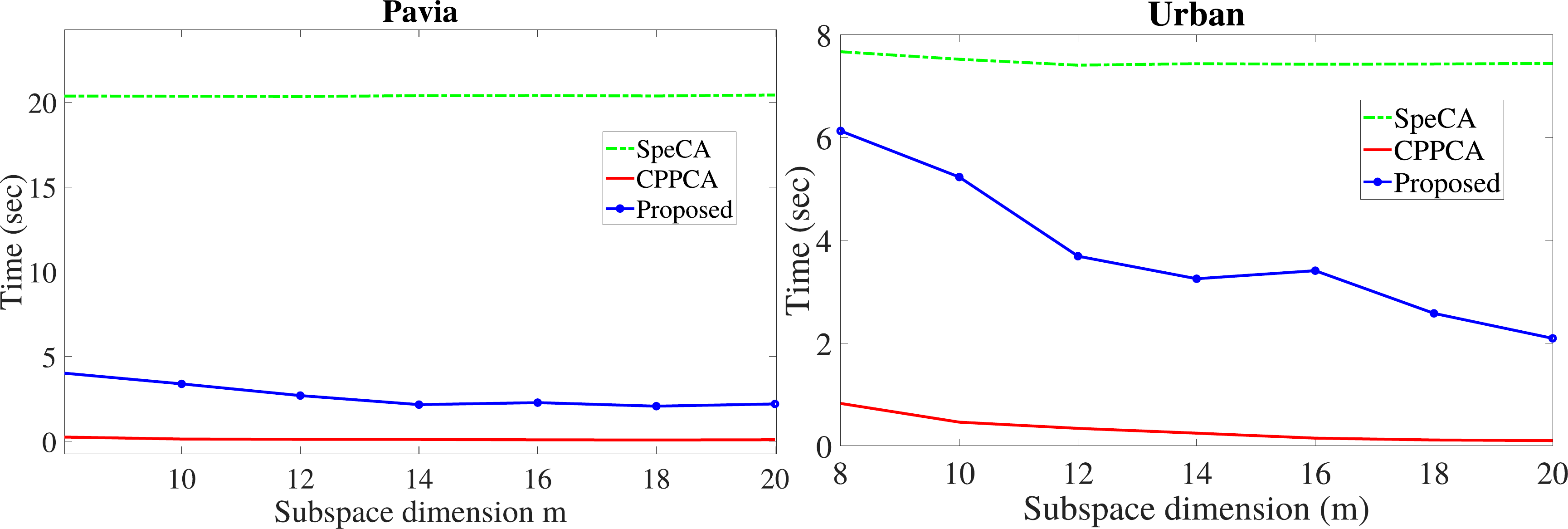}
    \caption{Comparison of the average execution time for SpeCA, CPPCA and the proposed algorithm. Note that the partition number is set to $p=l^2/m^2$ resulting in a reduction of the execution time when $m$ increases since the number of partitions decreases. }
    \label{fig:timecomparison}
\end{figure}

\begin{figure*}
    \centering
    \includegraphics[width=0.9\textwidth]{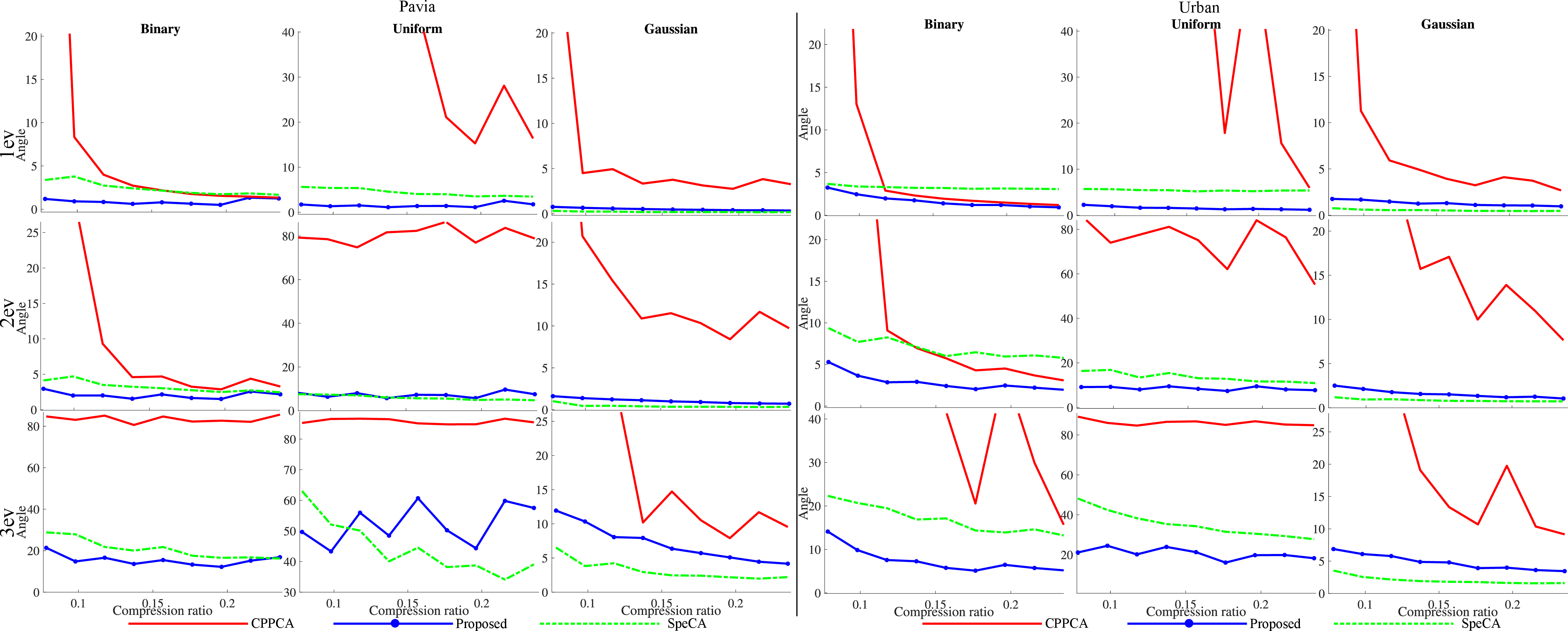}
    \caption{Average angle error of recovered eigenvectors with different compression ratios for Pavia (left), and Urban(right) images with different sensing matrices. Rows show the angle gap of the first, second, and third eigenvector.}
    \label{fig:eigenvectors}
\end{figure*}

Fig. \ref{fig:eigenvectors} shows the angle gap obtained with the two different images for the three different types of random projections. Results in Fig. \ref{fig:eigenvectors} are generated by running 20 times the proposed algorithm, along with CPPCA and SpeCA. The angles of the recovered eigenvectors are averaged. The  sensing protocol for the SpeCA algorithm is defined as $\mathbf{Y}_a = \mathbf{AX} \in \mathbb{R}^{m_a\times n}$, with $\mathbf{A} \in \mathbb{R}^{m_a\times l}$, $\mathbf{Y_b}= [\mathbf{B}_1\mathbf{x}_1,\mathbf{B}_2\mathbf{x}_2,\cdots,\mathbf{B}_n\mathbf{x}_n]$ and $\mathbf{B}_i \in \mathbb{R}^{m_b \times l}$, we set to $m_a=m-1$ and $m_b=1$. For these simulations, the signal was corrupted with additive Gaussian noise as in \eqref{eq:randomnoise} to yield 20 dB of SNR. The results show that the angle gap of the recovered eigenvectors is less when the proposed algorithm is used with any type of projection matrix. Note that SpeCA produces similar results to the proposed algorithm when Gaussian projection matrices are used. However, the proposed algorithm outperforms SpeCA when Binary and Uniform matrices are used. Additionally, Figure \ref{fig:timecomparison} shows the running time for the three algorithms by varying the dimension of the subspace $m$. For the proposed method the stopping criterium was set to be the relative tolerance given by $\|\bm{\Sigma}_k - \bm{\Sigma}_{k-1} \|/\| \bm{\Sigma}_k \|\leq 1e^{-4}$. It can be seen that SpeCA requires 37 seconds for Pavia in contrast to 0.6 and 0.2 seconds for proposed and CPPCA, respectively. Even though CPPCA is the fastest method, the reconstruction quality is up to two orders of magnitude worst, as shown in Fig. \ref{fig:eigenvectors3}. Note that the number of partitions for CPPCA and the proposed method is chosen following Lemma \eqref{lemma:singular}; when the dimension $m$ increases, the number of partitions $p$ decreases reducing the computation time.
\vspace*{-0.3cm}
\subsection{Error term and filtered gradient analysis}
\begin{figure}[H]
    \centering
    \includegraphics[scale=0.30]{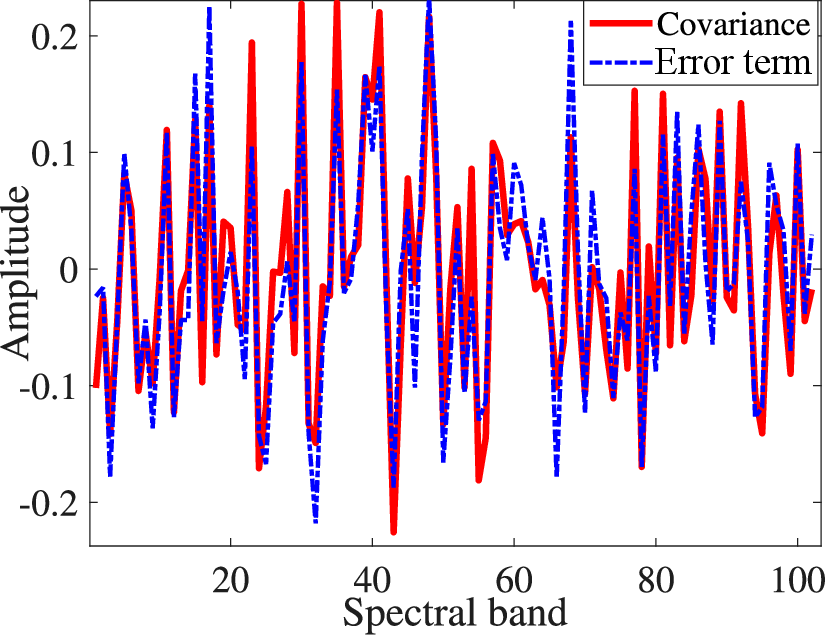}
     \includegraphics[scale=0.30]{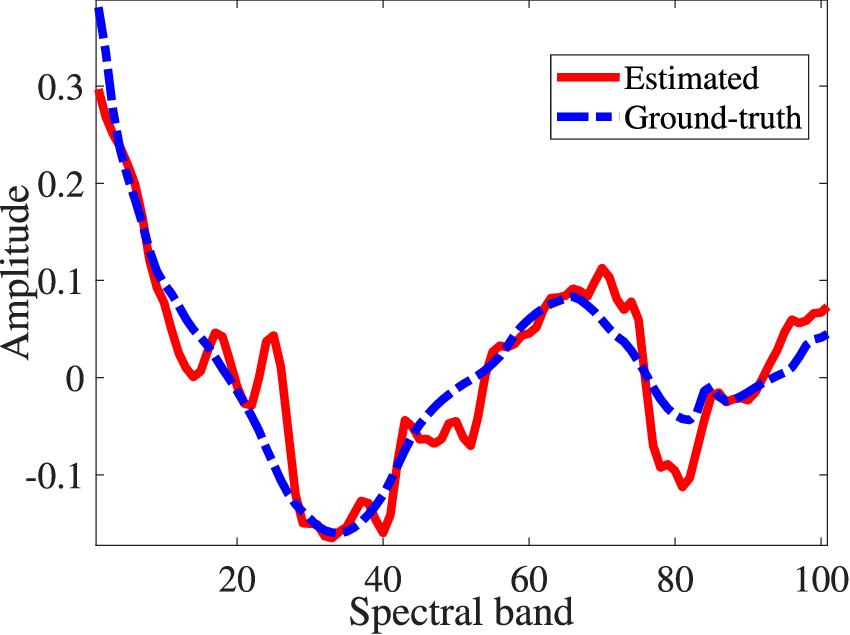}
    \caption{Comparison of the fourth eigenvector of the estimated covariance matrix with: (left) non-filtered gradient and the first eigenvector of the error term. (right) Filtered gradient and the fourth eigenvector of the truth covariance matrix. }
    \label{fig:biascov}
\end{figure}

In this section, the error term is numerically analyzed. For test purposes, we assume that the truth covariance is known so that the error matrices $\mathbf{R}_i$ are computed as $\mathbf{R}_i=\mathbf{S}-\mathbf{S}_i$, and the error is calculated as in \eqref{eq:bias}. The covariance matrix is estimated using the proposed algorithm without filtering the gradient, and its eigenvectors are compared with the error term $\mathbf{B}$. This is because when no filtering is applied, we observe in the simulations that some eigenvectors are corrupted with high-frequency noise. Fig. \ref{fig:biascov} (left) shows the eigenvector's visual comparison when no filter is applied on the gradient and an eigenvector of the bias term \eqref{eq:bias}. It can be seen that the fourth eigenvector of the recovered covariance matrix converges to the fourth eigenvector of $\mathbf{B}$, which computationally validates the statement in Lemma \ref{lema:bias}.

However, when the filtering procedure is applied using a Gaussian filter with $\sigma=1$, this corrupted eigenvector converges to the actual one; this is shown in Fig. \ref{fig:biascov} (right). Further analysis is shown in Appendix \ref{sec:errortermanalysis}. 
\vspace*{-0.4cm}
\subsection{Image reconstruction}
\begin{figure}[H]
    \centering
    \includegraphics[scale=0.27]{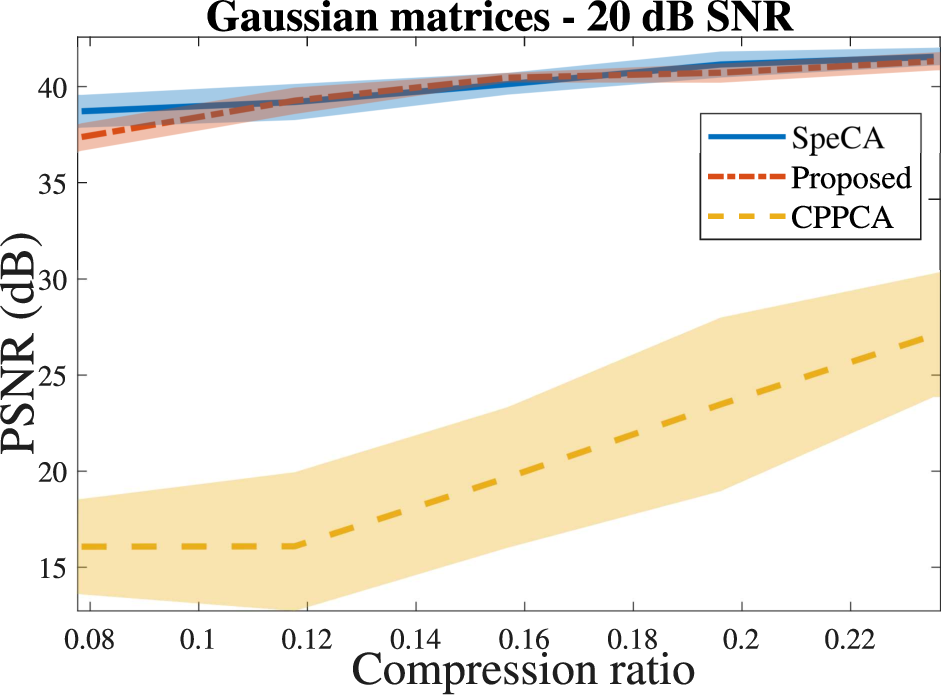}
    \includegraphics[scale=0.27]{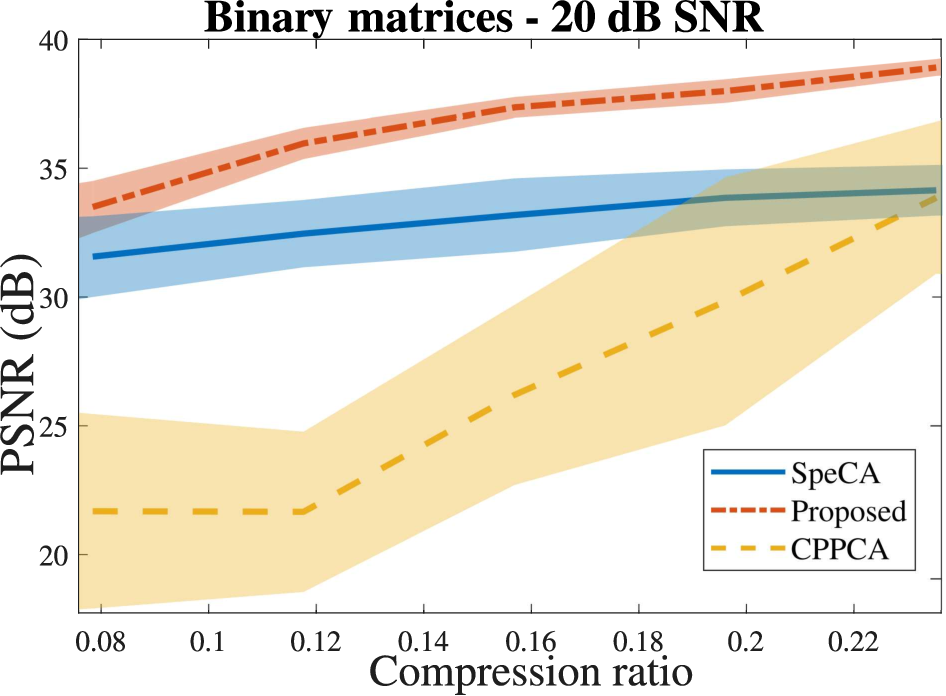}
    \caption{Comparison of the different reconstruction methods in terms of PSNR as a function of the compression ratio. The shaded areas represent the confidence interval.}
    \label{fig:reconsp}
\end{figure}

The underlying signal is recovered with the estimated eigenvectors using the method described in \cite{fowler2009compressive}. In particular, given the matrix $\mathbf{W}_m \in \mathbb{R}^{l\times m}$ containing $m$ recovered eigenvectors, the signal is estimated as
\begin{equation}
    \mathbf{X} = \mathbf{W}_m(\mathbf{P}^T\mathbf{W}_m)^{\dagger} \mathbf{Y},
    \label{eq:res11}
\end{equation}
where $\dagger$ is the Moore-Penrose inverse. Using this approach, the image is reconstructed, and the performance is compared against SpeCA and CPPCA algorithms. Figure \ref{fig:reconsp} shows the results for the Pavia centre image using the PSNR as a quality measurement.
It can be seen that using the estimator given in \eqref{eq:res11}, the proposed method outperforms both state-of-art counterparts, specially CPPCA, which exhibits a large dispersion on the performance. Note that, SpeCA results are similar to those obtained when Gaussian matrices are used, but the proposed method outperforms by up to 5 dB using binary matrices. 

\subsection{Optical implementation on DD-CASSI architecture}

\begin{figure}
    \centering
    \includegraphics[width=0.48\textwidth]{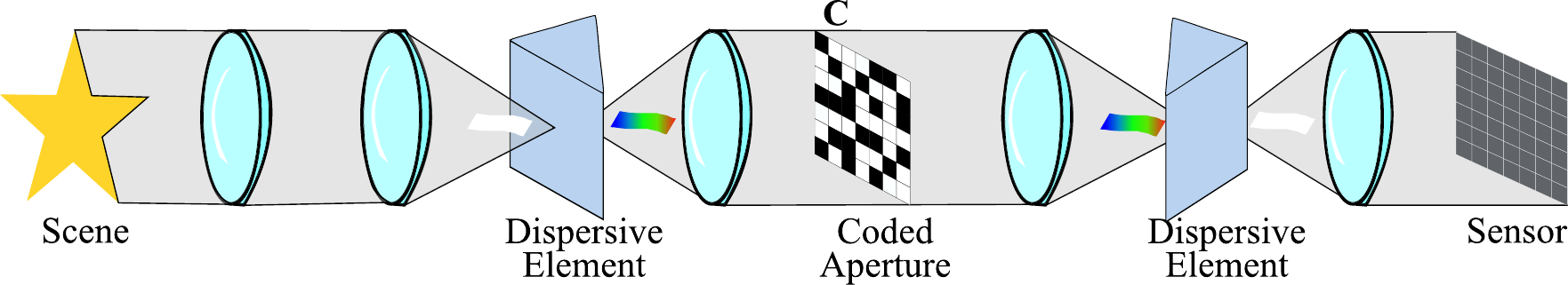}
    \caption{Schematic of DD-CASSI architecture.}
    \label{fig:ddcasi}
\end{figure}

Many implementable optical architectures model the sensing process as the vector formulation $\mathbf{y}=\mathbf{Hx}+\mathbf{n}$\cite{Correa2016, Gehm2007}, where $\mathbf{x}=\text{vec}(\mathbf{X})$ and $\mathbf{H}$ is the sensing matrix. However, the proposed method requires dividing the sensing problem into multiple independent sub-problems and expressing them in matrix form. This partition can be achieved in architectures like DD-CASSI\cite{Gehm2007} or SSCSI\cite{Lin2014} since they preserve the spatial independence in the sensor, i.e., the codification/compression occurs only along the spectral dimension. To convert the vector problem into the multiples matrix sub-problems, note that the sensing problem in DD-CASSI can be expressed as
\begin{eqnarray}\label{Eq:sensing2}
\mathbf{y}=
\begin{bmatrix}
\mathbf{y}_1\\
\mathbf{y}_2\\
\vdots\\
\mathbf{y}_n
\end{bmatrix}=
\begin{bmatrix}
\mathbf{P}^T_1 & 0 & \ldots & 0\\
0 & \mathbf{P}^T_2 & \ldots & 0\\
 \vdots & \dots & \dots & \vdots\\
0 & 0 & \ldots & \mathbf{P}^T_n\\
\end{bmatrix}
\begin{bmatrix}
\mathbf{x}_1\\
\mathbf{x}_2\\
\vdots\\
\mathbf{x}_n
\end{bmatrix}+\mathbf{r},
\end{eqnarray}
with $\mathbf{r}$ noise. From \eqref{Eq:sensing2} it can be seen that each pixel is coded by a different sensing matrix $\mathbf{P}_i$. In fact, \eqref{Eq:sensing2} is equivalent to \eqref{eq:partition} with $p=n$. Hence, if the number of sensing matrices is limited to $p<n$, \eqref{Eq:sensing2} can be re-written as \cite{HYCA}
\begin{eqnarray}\label{Eq:sensing3}
\mathbf{Y}=
\begin{bmatrix}
\mathbf{Y}_1\\
\mathbf{Y}_2\\
\vdots\\
\mathbf{Y}_{p}
\end{bmatrix}=
\begin{bmatrix}
\mathbf{P}^T_1 & 0 & \ldots & 0\\
0 & \mathbf{P}^T_2 & \ldots & 0\\
 \vdots & \dots & \dots & \vdots\\
0 & 0 & \ldots & \mathbf{P}^T_{p}\\
\end{bmatrix}
\begin{bmatrix}
\mathbf{X}_{1}\\
\mathbf{X}_{2}\\
\vdots\\
\mathbf{X}_{p}
\end{bmatrix}+\mathbf{E},
\end{eqnarray}
where $\mathbf{X}_i \in \mathbb{R}^{l\times n/p}$ is a matrix whose columns are the pixels coded by the same matrix $\mathbf{P}_i$, and $\mathbf{E}=[\mathbf{N}_1^T,\ldots,\mathbf{N}_p^T]^T$ is the noise. The schematic of the DD-CASSI, Figure \ref{fig:ddcasi}, shows the distribution of the optical elements. The sensing process consists of four main steps: first, the scene goes through a prism that induces a dispersion effect; second, the scene is modulated by a binary coded aperture $\mathbf{C}$; third, a second prism undo the dispersion of the first prism, and fourth the scene is integrated into the 2D sensor.

\begin{figure}[!htb]
    \centering
    \includegraphics[width=0.5\textwidth]{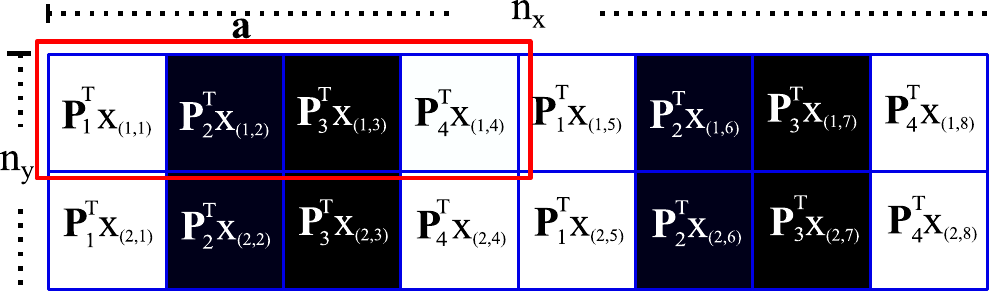}
    \caption{Sensing protocol design to limit the number of sensing matrices}
    \label{fig:sensingP}
\end{figure}

Limiting the number of sensing matrices $\mathbf{P}_i$ requires that the spatial distribution of the coded aperture $\mathbf{C}$ be designed to produce a limited number of code patterns in the spectral domain. One way to generate a limited number of sensing matrices consists of repeating a one-dimensional binary pattern $\mathbf{a}\in \mathbb{R}^{1\times p}$ along the spatial dimensions of the coded aperture $\mathbf{C}$. Specifically, the pattern $\mathbf{a}$ is repeated in each row of $\mathbf{C}$ in the same way; this concept is illustrated in Fig. \ref{fig:sensingP}. It can be seen that by repeating the pattern $\mathbf{a}$,  we can construct the matrix $\mathbf{X}_1=[\mathbf{x_{(1,1)}},\mathbf{x}_{(1,5)},\mathbf{x}_{(1,1)},\mathbf{x}_{(1,1)}]$ since they share the same matrix $\mathbf{P}_1$.

In a multishot setup, the spatial distribution of the coded aperture must change i.e., there is a $\mathbf{C}_t$ for each snapshot $t$. Nevertheless, to preserve the subsets distribution, only the entries of $\mathbf{a}_t$ change at each instant $t$. Hence, the coded aperture spatial distribution follows
\begin{equation}
\label{Eq:CodedAperture}
\text{vec}(\mathbf{C}_t)=
\left[\mathbf{1}_{n_y}\otimes(\mathbf{1}_{N_x/p}\otimes\mathbf{a}_{t})\right].
\end{equation}

\begin{figure}[!htb]
	\centering
		\includegraphics[width=0.35\textwidth]{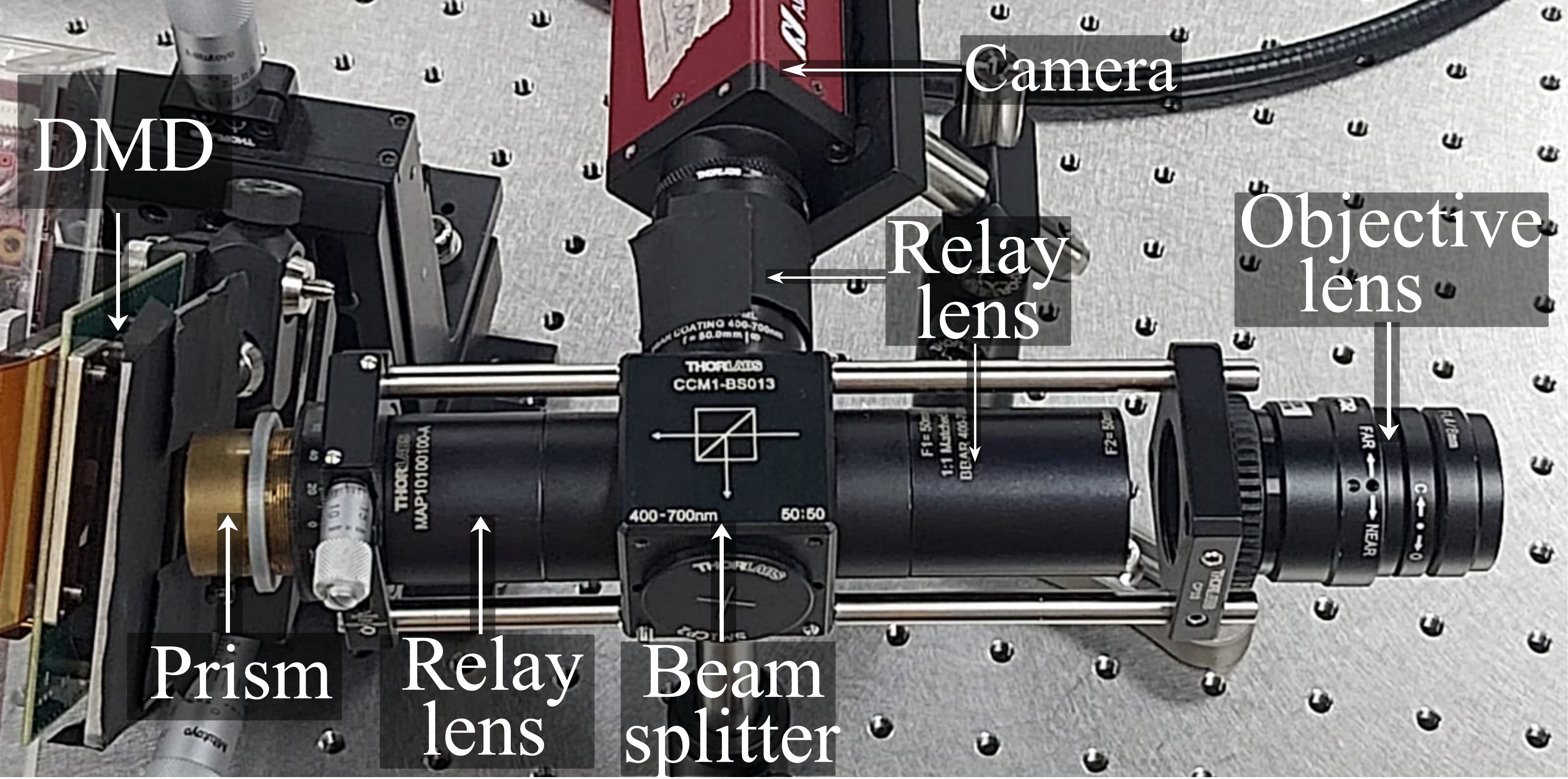}
	\caption{\label{fig:DD-arch}{\color{black} Optical implementation of the DD-CASSI architecture. }}
\end{figure}

We built a testbed in our laboratory as a proof-of-concept prototype based on\cite{marquez1,marquez2,marquez3}; the optical setup is shown in Fig. \ref{fig:DD-arch}. This optical device is made out of a Navitar lens ($12 mm$ FixedFocal Length, MVL12M23 - $12 mm$ EFL, $f/1.4$) as the objective lens to image the scene onto the image plane of a matched achromatic doublet pair (Thorlabs MAP10100100-A, $f1=100.0 mm, f2=100.0 mm$) to propagate the incoming wavefront through a beam splitter until to a second matched achromatic doublet pair relay lens (Thorlabs MAP10100100-A, $f1=100.0 mm, f2=100.0 mm$). This second relay lens transmits the wavefront through a double Amici prism coupled to a rotation mount (Thorlabs CRM1P, $30 mm$ cage rotation mount, \O 1") to image a dispersed version of the scene onto the digital micromirror device (DMD, Texas Instruments, D4120). Taking advantage of the DMD's mirror surface, the now dispersed-modulated wavefront is returned through to the prism until the L2 lens, where the prism undoes the dispersion effect. The resulting dispersed-coded-dispersed wavefront propagates through BS until a third matched achromatic doublet pair relay lens (L3) (Thorlabs MAP105050-A, $f1=50.0 mm, f2=50.0 mm$). Finally, the L3 lens focuses the dispersed-coded-dispersed wavefront onto the sensor (Stingray F-080B, $4.65 \mu m$ pixel size).

\begin{figure}
    \centering
    \includegraphics[width=0.46\textwidth]{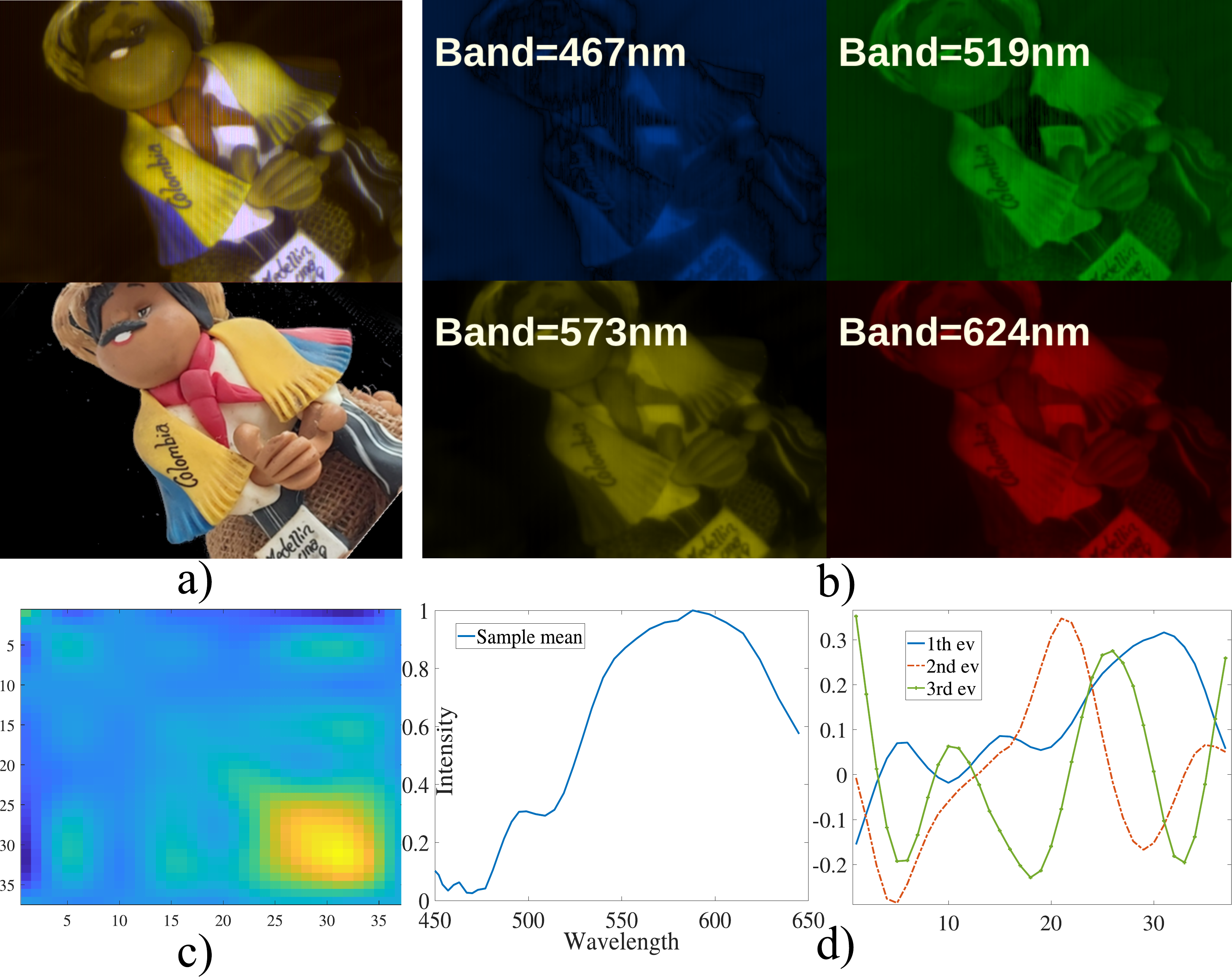}
    \caption{{\color{black}Reconstruction of the covariance matrix and hyperspectral image from data captured in our lab using the DD-CASSI optical architecture. a)-top RGB composite of the reconstructed hyperspectral image, a)-bottom RGB image acquired with a commercial camera. b) four out of the 37 spectral bands of the reconstructed hyperspectral image. c) Recovered covariance matrix. d) Sample mean and first three eigenvectors of the covariance matrix.}}
    \label{fig:real}
\end{figure}

The coded aperture was designed using \eqref{Eq:CodedAperture} to produce a limited number of patterns with $m=8$ snapshots. We placed the prism in a distance such that the dispersion generated $l=37$ spectral bands in the sensor, and the spatial resolution of the scene was $356\times 512$ pixels; this setup achieves a $79\%$ of compression of the image. Based on Lemma \eqref{lemma:singular}, the optimal number of partitions must be $p>l^2/m^2=21.4$, so we generated 24 partitions. The sample mean was computed using \eqref{eq:mean} and subtracted from the measurements. Additionally, we used $\tau=2e^{-4}$ and 200 iterations in the covariance recovery algorithm. We took the first five eigenvectors for the image reconstruction, and used them in \eqref{eq:res11}. Overall, the whole process, including covariance matrix recovery and image reconstruction, took 0.85 seconds on average. Figure \ref{fig:real} shows an RGB composite of the hyperspectral image reconstructed and the RGB image captured with a commercial camera for comparison purposes. Fig. \ref{fig:real}-b) shows four out of the 37 reconstructed spectral bands; these 37 spectral bands are in the range of 450 nm to 650 nm with a spectral resolution going from $2$nm in blue spectral bands until $10$nm in the red spectral bands. Figure \ref{fig:real}-c) shows the recovered covariance matrix; Fig. \ref{fig:real}-d) shows the sample mean and the three eigenvectors associated with the largest eigenvalues. The figure shows that the RGB composite resembles the colors obtained with the commercial camera's high-resolution camera, which gives insights into the correct reconstruction.

\section{Discussion} \label{sec:discussion}
 One limitation of the proposed method is that it requires multi-shot acquisition to correct covariance reconstruction. Specifically, in the reconstruction step, when a single shot is acquired, inverse problem \eqref{eq:res11} produces a rank-one solution that is not accurate. However, nowadays, cameras can acquire shots at a high-speed rate, reducing the impact of these limitations. Additionally, prior knowledge of the covariance matrix is required to set the convex set (e.g., Low-rank or Toeplitz). {\color{black} On the other hand, the partition of the data makes the method impractical to work with few realizations compared to the number of spectral bands. That is because the sample covariances matrices $\mathbf{S}_i$ will be poor estimators, and the error associated with the partition will increase dramatically. Nevertheless,  the number of pixels is much greater than the number of spectral bands in imaging applications. }
\vspace*{-0.2cm}
\section{Conclusion}\label{sec:conclusions}
We proposed an algorithm to recover the covariance matrix from a set of compressive measurements using a strategy-based projection onto convex subsets. The algorithm is based on the projected gradient method. The theoretical results show that although the splitting procedure induces an error term, it can be mitigated using a filtered gradient. Additionally, this error is proportional to the number of partitions; nevertheless, more partitions improve the condition of the information matrix; thus, choosing the correct number of partitions is critical. For that reason, a lower bound for the optimal number of partitions is proposed. Experimental results show that the proposed method outperforms state-of-art algorithms CPPCA and SpeCA. The experiments were performed using two different hyperspectral images, for which the proposed method attained better results in terms of MSE and angle GAP, which translates in a gain of up to 10 dB of PSNR in comparison with CPPCA and up to 4 dB PSNR concerning SpeCA. Additionally, the algorithm was tested with real data from the laboratory using DD-CASSI architecture. It can be seen that the reconstruction process is fast and robust since the RGB composite resembles the RGB image of the scene.

\ifCLASSOPTIONcaptionsoff
  \newpage
\fi



%\bibliographystyle{IEEEtran}

\bibliography{main}
\bibliographystyle{ieeetr}

\end{document}

\setcounter{equation}{0}
\setcounter{figure}{0}
\renewcommand{\theequation}{S\arabic{equation}}
\renewcommand{\thefigure}{S\arabic{figure}}
\title{Supplementary material: \\ Covariance Estimation from Compressive Data Partitions using a Projected Gradient-based Algorithm}

\author{Jonathan Monsalve,
        Juan Ramirez,
        I\~naki Esnaola,~\IEEEmembership{Member,~IEEE}
        Henry Arguello,~\IEEEmembership{Senior Member,~IEEE \vspace{-8mm}}% <-this % stops a space

\thanks{The work of J. Monsalve was supported by the Colciencias/Department of Santander Scholarship (771 of 2016). J. Monsalve is with the Department of Electrical Engineering, Universidad Industrial de Santander, Bucaramanga 680002, Colombia. J. Ramirez is with the Department of Computer Science, Universidad Rey Juan Carlos, Madrid, 28933, Spain. I. Esnaola is with Department of Automatic Control and Systems Engineering, The University of Sheffield, Western Bank, Sheffield, UK. H. Arguello is with the Department Systems Engineering and Informatics, Universidad Industrial de Santander, Bucaramanga 680002, Colombia(e-mail: henarfu@uis.edu.co)}% <-this % stops a space
%\thanks{Manuscript accepted 01 Feb. 2020; revised xxxx xx, xxxx.}
}
%\thanks{Manuscript received xxxx xx, xxxx; revised xxxx xx, xxxx.}}

\markboth{IEEE Transactions on Image Processing, 2021}%
{Shell \MakeLowercase{\textit{et al.}}: Bare Demo of IEEEtran.cls for IEEE Journals}

\maketitle

\IEEEpeerreviewmaketitle

% or


\appendices

\vspace{-0.15cm}
\section{Proof: The expected value of the error term is zero.}\label{sec:proofR}
The proposed method assumes the covariance matrix of each subset $\mathbf{X}_i$ is 
\begin{equation}
    \mathbf{S}_i = \mathbf{S} + \mathbf{R}_i,
\end{equation}
where $\mathbf{R}_i$ is the error of the subset covariance matrix estimate. Note that 
\begin{equation}
    \begin{aligned}
    \mathbf{S} = \frac{1}{n} \sum_{j=1}^{n} \mathbf{x}_j\mathbf{x}_j^T =& \frac{1}{n} \Big(\sum_{j_1 \in S_1}  \mathbf{x}_{j_1}\mathbf{x}_{j_1}^T + \sum_{j_2 \in S_2}  \mathbf{x}_{j_2}\mathbf{x}_{j_2}^T + \\ & \cdots + \sum_{j_p \in S_p}  \mathbf{x}_{j_p}\mathbf{x}_{j_p}^T \Big),
    \label{eq:39}
    \end{aligned}
\end{equation}
\[ = \frac{1}{n} (\mathbf{X}_1\mathbf{X}_1^T + \mathbf{X}_2\mathbf{X}_2^T + \cdots + \mathbf{X}_p\mathbf{X}_p^T )\]
which yields to
\begin{equation}
    \begin{aligned}
    \mathbf{S} = &  \frac{1}{p}((\mathbf{S} + \mathbf{R}_1) +(\mathbf{S} + \mathbf{R}_2) + \cdots + (\mathbf{S} + \mathbf{R}_p) )
    \end{aligned}
    \label{eq:sumcov1}
\end{equation}
Computing the expectation in both sides yields to
\begin{equation}
    \begin{aligned}
    \mathbb{E}[\mathbf{S}] = & \mathbb{E}[\mathbf{S}] + \frac{1}{p} \mathbb{E}[ \mathbf{R}_1 + \mathbf{R}_2 + \cdots + \mathbf{R}_p ],
    \end{aligned}
    \label{eq:sumcov2}
\end{equation}
which implies that $\mathbb{E}[ \mathbf{R}_1 + \mathbf{R}_2 + \cdots + \mathbf{R}_p] = 0$.

% proof lemma

\section{Proof of the lemma \ref{lemma:zero}}\label{sec:proofentries}

Let us define $\mathbf{H}_i \equiv \mathbf{P}_i\mathbf{P}_i^T = [\mathbf{h}_1,\mathbf{h}_2,\cdots,\mathbf{h}_l]$, and define the matrix
\begin{equation}
    \mathbf{B}_i \equiv \mathbf{H}_i\mathbf{R}\mathbf{H}_i,
\end{equation}
where $\mathbf{H}_i$ is symmetric, and $\mathbb{E}[\mathbf{R}]=0$ (proof in appendix \ref{sec:proofR}). For simplicity, the $i$ index of matrices $\mathbf{B}$ and $\mathbf{R}$ is dropped. The expected value of the entries of the matrix $\mathbf{B}_{k,j} = \mathbf{h}_k^T\mathbf{R}\mathbf{h}_j$ is given by
\begin{equation}
    \begin{aligned}
    \mathbb{E}[\mathbf{h}_k^T\mathbf{R}\mathbf{h}_j] &= \mathbb{E}[\text{tr}(\mathbf{h}_k^T\mathbf{R}\mathbf{h}_j)] = \mathbb{E}[\text{tr}(\mathbf{R}\mathbf{h}_j \mathbf{h}_k^T)]  \\
    &= \text{tr}(\mathbb{E}[\mathbf{R}\mathbf{h}_j \mathbf{h}_k^T)].
    \end{aligned}
\end{equation}
Additionally, the matrix $\mathbf{h}_j\mathbf{h}_k$ is deterministic,
and $\mathbb{E}[\mathbf{R}]=0$, hence
\begin{equation}
    \begin{aligned}
    \mathbb{E}[\mathbf{h}_k^T\mathbf{R}\mathbf{h}_j] &= 0.
    \end{aligned}
\end{equation}

\section{Error term of the proposed gradient method}\label{sec:errorproof}

 The gradient of $f(\bm{\Sigma})$ is given in \eqref{eq:grad1}. However, as mentioned above, the covariance matrices are not equal, thus \eqref{eq:grad1} is rewritten as
\begin{equation}
    \nabla \tilde{f}(\bm{\Sigma}) = \sum_{i=1}^{p} \mathbf{P}_i(\bm{\tilde{\Sigma}}_i-\mathbf{P}_i^T\mathbf{S}_i\mathbf{P}_i)\mathbf{P}_i^T + \tau \nabla \psi(\bm{\Sigma}).
    \label{eq:grad2}
\end{equation}
Plugging \eqref{eq:error} into \eqref{eq:grad2} and assuming that $\mathbf{S}=\bm{\Sigma}$ we have that
\begin{equation}
    \begin{aligned}
    \nabla \tilde{f}(\bm{\Sigma}) &= \sum_{i=1}^{p} \mathbf{P}_i(\bm{\tilde{\Sigma}}_i-\mathbf{P}_i^T(\mathbf{S}+\mathbf{R}_i)\mathbf{P}_i)\mathbf{P}_i^T + \tau \nabla \psi(\bm{\Sigma})\\& = \sum_{i=1}^{p} \mathbf{P}_i(\bm{\tilde{\Sigma}}_i-\mathbf{P}_i^T(\bm{\Sigma}+\mathbf{R}_i)\mathbf{P}_i)\mathbf{P}_i^T + \tau \nabla \psi(\bm{\Sigma}).
     \end{aligned}
    \label{eq:28}
\end{equation}
After some algebraic operations, \eqref{eq:28} can be rewritten as
\begin{equation}
    \begin{aligned}
    \nabla \tilde{f}(\bm{\Sigma}) =& \sum_{i=1}^{p} \mathbf{P}_i(\bm{\tilde{\Sigma}}_i-\mathbf{P}_i^T\bm{\Sigma}\mathbf{P}_i)\mathbf{P}_i^T - \sum_{i=1}^p \mathbf{P}_i\mathbf{P}_i^T\mathbf{R}_i\mathbf{P}_i\mathbf{P}_i^T \\+& \tau \nabla \psi(\bm{\Sigma}).
    \end{aligned}
    \label{eq:gradbiased}
\end{equation}
Comparing \eqref{eq:grad} and \eqref{eq:gradbiased}, we can see that 
\begin{equation}
    \nabla \tilde{f}(\bm{\Sigma}) = \nabla f(\bm{\Sigma}) -\sum_{i=1}^p \mathbf{P}_i\mathbf{P}_i^T\mathbf{R}_i\mathbf{P}_i\mathbf{P}_i^T.
\end{equation}
Hence, the error term of the gradient induced by the splitting procedure is
\begin{equation}
     Error[\nabla \tilde{f}(\bm{\Sigma})]=-\sum_{i=1}^p \mathbf{P}_i\mathbf{P}_i^T\mathbf{R}_i\mathbf{P}_i\mathbf{P}_i^T.
    \label{eq:bias}
\end{equation}

\section{Proof of the Cram\'er-Rao Lower Bound for the estimator}\label{sec:cramer}
The variance of the estimator is bounded by
\begin{equation}
     \text{var}(\bm{\tilde{\Sigma}})\geq\text{tr}(\mathbf{I}(\bm{\Sigma})^{-1})
\end{equation}
where $\mathbf{I}(\bm{\Sigma})$ is the fisher information matrix. To compute $\mathbf{I}(\bm{\Sigma})$ we observe that
\begin{equation}
    \mathbf{\tilde{S}}_i=\frac{n}{p}\mathbf{Y}_i \mathbf{Y}_i^T,
    \label{eq:quadmes}
\end{equation}
with $\frac{n}{p}\mathbf{\tilde{S}}_i \sim \mathcal{W}(\mathbf{P}_i^T\bm{\Sigma}\mathbf{P}_i + \bm{\Sigma}_N, n/p)$, and set $r = n/p$ so that that each subset contains $n/p$ samples. The likelihood function is given by
\begin{equation}
    \begin{aligned}
        f(\mathbf{\tilde{S}}_1,\cdots,\mathbf{\tilde{S}}_p| \bm{\Sigma}) \propto &\prod_{i=1}^p \frac{1}{|\bm{\Sigma}_N + \mathbf{P}_i^T\bm{\Sigma}\mathbf{P}_i|^{\frac{n}{p}}} \times \left| \mathbf{Y}_i\mathbf{Y}_i^T \right|^{\frac{n}{p}-m}\\ &\times \text{etr}\left\{- (\bm{\Sigma}_N + \mathbf{P}_i^T\bm{\Sigma}\mathbf{P}_i)^{-1} (\mathbf{Y}_i\mathbf{Y}_i^T)  \right\},
    \end{aligned}
\end{equation}
where $|.|$ stands for the determinant and $\text{etr}\{.\}$ the exponential of the trace. Applying the logarithm in both sides yields
\begin{equation}
        \begin{aligned}
        &F(\mathbf{\tilde{S}}_1,\cdots,\mathbf{\tilde{S}}_p| \bm{\Sigma}) \propto \frac{-n}{p}\sum_{i=1}^p \log{|\bm{\Sigma}_N + \mathbf{P}_i^T\bm{\Sigma}\mathbf{P}_i|}+\\& (\frac{n}{p}-m)\sum_{i=1}^p \log{|\mathbf{Y}_i\mathbf{Y}_i^T|} - \sum_{i=1}^p \text{tr}\left((\bm{\Sigma}_N + \mathbf{P}_i^T\bm{\Sigma}\mathbf{P}_i)^{-1} (\mathbf{Y}_i\mathbf{Y}_i^T)\right),
        \end{aligned}
\end{equation}
where $F(\mathbf{\tilde{S}}_1,\cdots,\mathbf{\tilde{S}}_p| \bm{\Sigma}) = \log{f(\mathbf{\tilde{S}}_1,\cdots,\mathbf{\tilde{S}}_p| \bm{\Sigma})}$. Differentiating twice with respect to $\bm{\sigma}=\text{vec}(\bm{\Sigma})$ yields 
\begin{equation}
    \begin{aligned}
    &\frac{\partial^2 F(\mathbf{\tilde{S}}_i|\bm{\Sigma})}{\partial \bm{\sigma} \partial \bm{\sigma}^T} = \frac{n}{p}\sum_{i=1}^p \mathbf{P}_i\mathbf{A}_i^T\mathbf{P}_i^T\otimes \mathbf{P}_i\mathbf{A}_i\mathbf{P}_i^T + \\& \mathbf{P}_i \mathbf{A}_i^T \mathbf{\tilde{S}} \mathbf{A}_i^T\mathbf{P}_i^T\otimes  \mathbf{P}_i \mathbf{A}_i\mathbf{P}_i^T - \mathbf{P}_i \mathbf{A}_i^T\mathbf{P}_i^T \otimes \mathbf{P}_i \mathbf{A}_i \mathbf{\tilde{S}} \mathbf{A}_i\mathbf{P}_i^T,
    \end{aligned}
    \label{eq:fisher1}
\end{equation}
with $\mathbf{A}_i = (\bm{\Sigma}_N + \mathbf{P}_i^T\bm{\Sigma}\mathbf{P}_i)^{-1}$. The fisher information matrix is computed by calculating the expectation of \eqref{eq:fisher1} which yields to
\begin{equation}
    \mathbf{I}(\bm{\Sigma}) = \frac{n}{p}\sum_{i=1}^p \mathbf{P}_i\mathbf{A}_i^T\mathbf{P}_i^T\otimes \mathbf{P}_i\mathbf{A}_i\mathbf{P}_i^T.
    \label{eq:fisher2}
\end{equation}
Note that, from \eqref{eq:fisher1} to \eqref{eq:fisher2} we used $ \mathbb{E}\left[\mathbf{P}_i \mathbf{A}_i^T \mathbf{\tilde{S}} \mathbf{A}_i^T\mathbf{P}_i^T\otimes  \mathbf{P}_i \mathbf{A}_i\mathbf{P}_i^T\right] = \frac{n}{p} \mathbf{P}_i \mathbf{A}_i^T\mathbf{P}_i^T\otimes  \mathbf{P}_i \mathbf{A}_i\mathbf{P}_i^T$, since $\mathbf{P}_i, \mathbf{A}_i$ are deterministic matrices and $\mathbb{E}\left[ \mathbf{\tilde{S}} \right] = \frac{n}{p}(\bm{\Sigma}_N + \mathbf{P}_i^T\bm{\Sigma}\mathbf{P}_i)$. Hence, the estimator variance of $\bm{\Sigma}$ is bounded by
\begin{equation}
    \text{var}(\bm{\tilde{\Sigma}})\geq \frac{p}{n}\text{Tr}\left(\sum_{i=1}^p \mathbf{P}_i\mathbf{A}_i^T\mathbf{P}_i^T\otimes \mathbf{P}_i\mathbf{A}_i\mathbf{P}_i^T\right)^{-1}.
\end{equation}
\section{Proof lemma \ref{lemma:singular}}\label{sec:singular}
Given matrices $\mathbf{P}_i \in \mathbb{R}^{l\times m}$ and $\mathbf{A}_i \in \mathbb{R}^{m\times m}$ with $m<=l$ it holds that 
\begin{equation}
    \text{rank}(\mathbf{P}_i\mathbf{A}_i\mathbf{P}_i^T) = \text{rank}(\mathbf{P}_i\mathbf{A}_i^T\mathbf{P}_i^T)\leq m,
\end{equation}
using the fact that $\text{rank}(\mathbf{C}\otimes \mathbf{D})=\text{rank}(\mathbf{C}) \text{rank}(\mathbf{D})$ and $\text{rank}(\mathbf{C} + \mathbf{D})\leq\text{rank}(\mathbf{C}) + \text{rank}(\mathbf{D})$, it can be concluded that 
\begin{equation}
    \text{rank}\left(\sum_{i=1}^p \mathbf{P}_i\mathbf{A}_i^T\mathbf{P}_i^T\otimes \mathbf{P}_i\mathbf{A}_i\mathbf{P}_i^T\right)\leq m^2 p.
\end{equation}
Hence, for fixed values of $m$ and $l$ the matrix
\begin{equation}
    \mathbf{I}(\bm{\Sigma})=\sum_{i=1}^p \mathbf{P}_i\mathbf{A}_i^T\mathbf{P}_i^T\otimes \mathbf{P}_i\mathbf{A}_i\mathbf{P}_i^T \in \mathbb{R}^{l^2 \times l^2},
\end{equation}
is singular if $p<l^2/m^2$.
\section{Filtering reduce the variance of the error term}\label{sec:varred}
{\color{black}
For that, note that  $\mathbb{E}_{B}(\hat{f}(\bm{\Sigma})_{ij})=f(\bm{\Sigma})_{ij}$ with $\sigma_f = \text{var}(f(\bm{\Sigma})_{ij})=\text{var}(\mathbf{B}_{i,j})$ assuming that the entries of $\mathbf{B}$ are i.i.d.; then averaging in $k\times k$ window yields to
\begin{equation}
\begin{aligned}
    \mathbb{E}_{B}&\left[\mathbf{K} * \nabla \hat{f}(\bm{\Sigma})\right]_{i,j} = \frac{1}{k^2}\mathbb{E}\left[\sum_{\iota,\rho = -k/2 }^{k/2} \nabla \hat{f}(\bm{\Sigma})_{i+\iota,j+\rho}\right],
\end{aligned}
\label{eq:variance}
\end{equation}
with $\mathbb{E}_{B}$ the expectation over $\mathbf{B}$. Replacing \eqref{eq:gradbiased} into \eqref{eq:variance} 
\begin{equation}
\begin{aligned}
    \mathbb{E}\left[\mathbf{K} * \nabla \hat{f}(\bm{\Sigma})\right]_{i,j} &= \frac{1}{k^2}\sum_{\iota,\rho}  \mathbb{E}\left[\nabla f(\bm{\Sigma})_{i+\iota,j+\rho} + \mathbf{B}_{i+\iota,j-\rho}\right]\\
    &=\frac{1}{k^2}\sum_{\iota,\rho}  \nabla f(\bm{\Sigma})_{i+\iota,j+\rho}.
\end{aligned}
\label{eq:variance1}
\end{equation}
The variance is 
\begin{equation}
\begin{aligned}
&\mathbb{E}_{B}\left[\left\{\big( \mathbf{K} * \nabla \hat{f}(\bm{\Sigma})\big)_{i,j} - \mathbb{E}_{B}\left[\mathbf{K} * \nabla \hat{f}(\bm{\Sigma})\right]_{i,j}\right\}^2 \right] \\
&= \mathbb{E}\left[\left(\frac{1}{k^2}\sum_{\iota,\rho}  \mathbf{B}_{i+\iota,j-\rho}\right)^2\right]=\frac{\sigma_f}{k^2}.
\end{aligned}
\end{equation}
}
\section{Bound of the norm for the error term}\label{sec:bounderror}
{\color{black}

	    Defining $\mathbf{H}_i=\mathbf{P}_i\mathbf{P}_i^T$, the $\ell_2$ norm of the error term for a fixed $i$ in lemma \eqref{lema:bias} is given by 
	    \begin{equation}
	        \| \mathbf{H}_i\mathbf{R}_i \mathbf{H}_i\|_2 \leq \sigma_{max}(\mathbf{H}_i)\|\mathbf{R}_i\mathbf{H}_i\|_2\leq \sigma_{max}(\mathbf{H}_i)^2\|\mathbf{R}_i\|_2.
	    \end{equation}
	    From (10) we have that with probability at least $1 - 2\exp{(-t^2 l)}$ it holds that $\|\mathbf{R}_i\|\leq \epsilon$. Using this fact, with high probability it holds \begin{equation}
	        \| \mathbf{H}_i\mathbf{R}_i \mathbf{H}_i\|_2 \leq \sigma_{max}(\mathbf{H}_i)^2 \epsilon \leq \sigma_m^2 \epsilon,
	        \label{eq:ineq1}
	    \end{equation}
	    with $\sigma_m = max(\sigma_{max}(\mathbf{H}_1)^2,\ldots, \sigma_{max}(\mathbf{H}_p)^2)$, and the variance is bounded by $\| \sum_{i=1}^p \mathbb{E}(\mathbf{H}_i\mathbf{R}_i \mathbf{H}_i)^2 \|<\sigma_{H}$. Hence, the Non-commutative Bernstein-type inequality\cite{Tropp_2011} establishes that for every $t\geq 0$
	    \begin{equation}
	        \mathbb{P}\left\{ \left\| \sum_{i=1}^{p} \mathbf{H}_i\mathbf{R}_i \mathbf{H}_i\right\|_2 \geq t \right\} \leq 2\times l\times  e^{\frac{-t^2/2}{\sigma_H^2 + \sigma_m^2 \epsilon t/3}}.
	    \end{equation}

	    Additionally, using the triangle inequality and \eqref{eq:ineq1}, yields to
	    \begin{equation}
	        \left \|\sum_{i=1}^p \mathbf{H}_i\mathbf{R}_i \mathbf{H}_i\right \|_2 \leq p \sigma_{m}^2 \epsilon,
	        \label{eq:bound}
	    \end{equation}
	    with high probability. Equation \eqref{eq:bound} shows that the error term increases linearly with the number of partitions, but additionally more partitions also increase the error term $\epsilon$ which increases in a quadratic manner.
}
\section{Experimental distribution of the error term}
\begin{figure}
    \centering
    \includegraphics[width=0.48\textwidth]{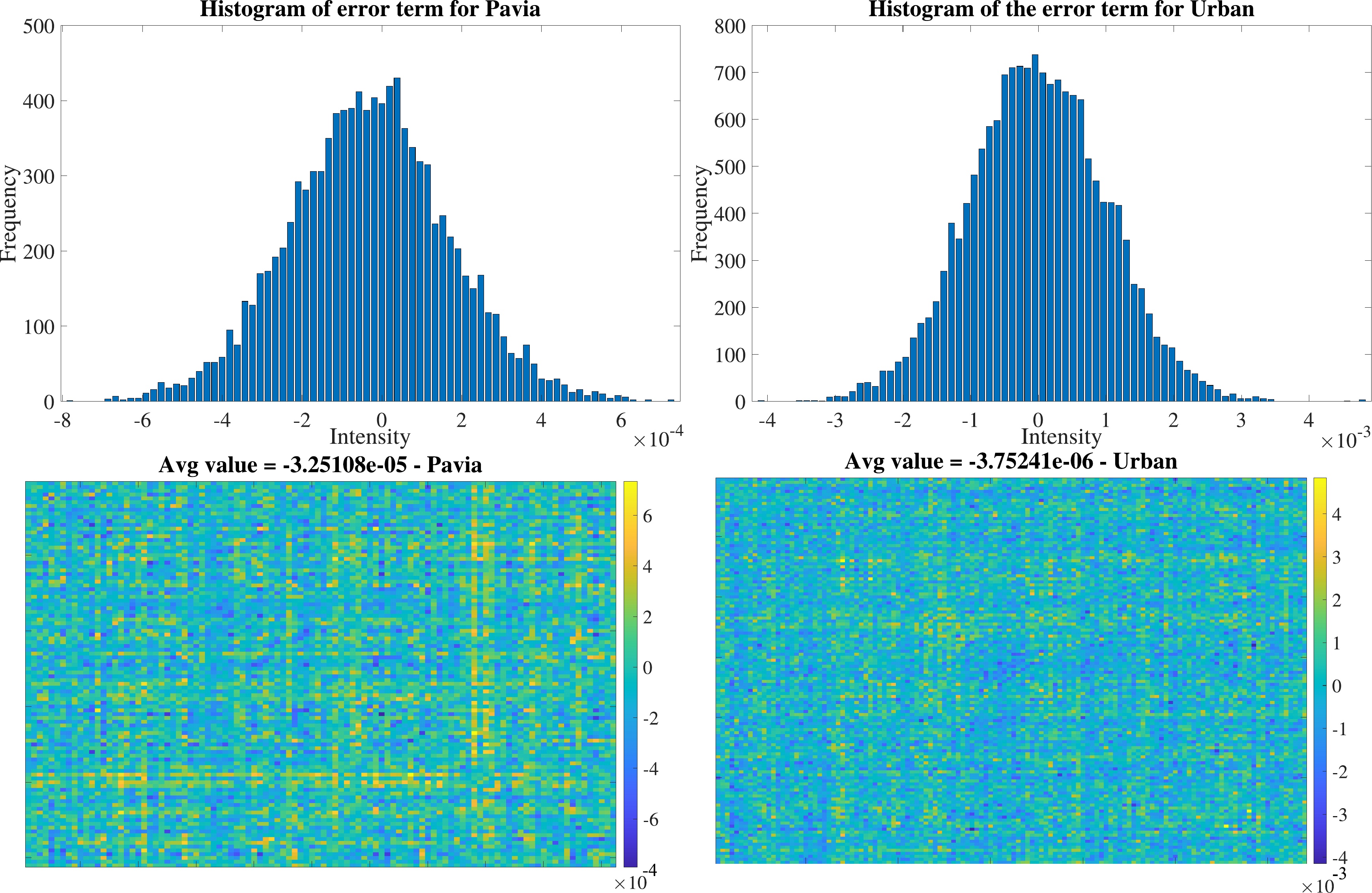}
    \caption{{\color{black}Error term of the partition procedure given in \eqref{eq:error}. Top: Histogram of the error term matrix entries. Bottom: Spatial distribution of the error term.}}
    \label{fig:errortermhist}
\end{figure}

{\color{black} On the other hand, lemma \ref{lemma:zero} states that the expected value of each entry of the error term is zero, in order to validate it computationally this term \eqref{eq:error} is computed. Figure \ref{fig:errortermhist}-bottom shows the spatial distribution of the error term for both images. Figure \ref{fig:errortermhist}-top shows the histogram of the error term values. It can be seen that the values are placed around zero and the sample mean is $-3.25e^{-5}$, $-3.75e^{-6}$ for Pavia and Urban respectively. }

\section{Error term analysis}\label{sec:errortermanalysis}
\begin{figure}[!htb]
    \centering
    \includegraphics[scale=0.31]{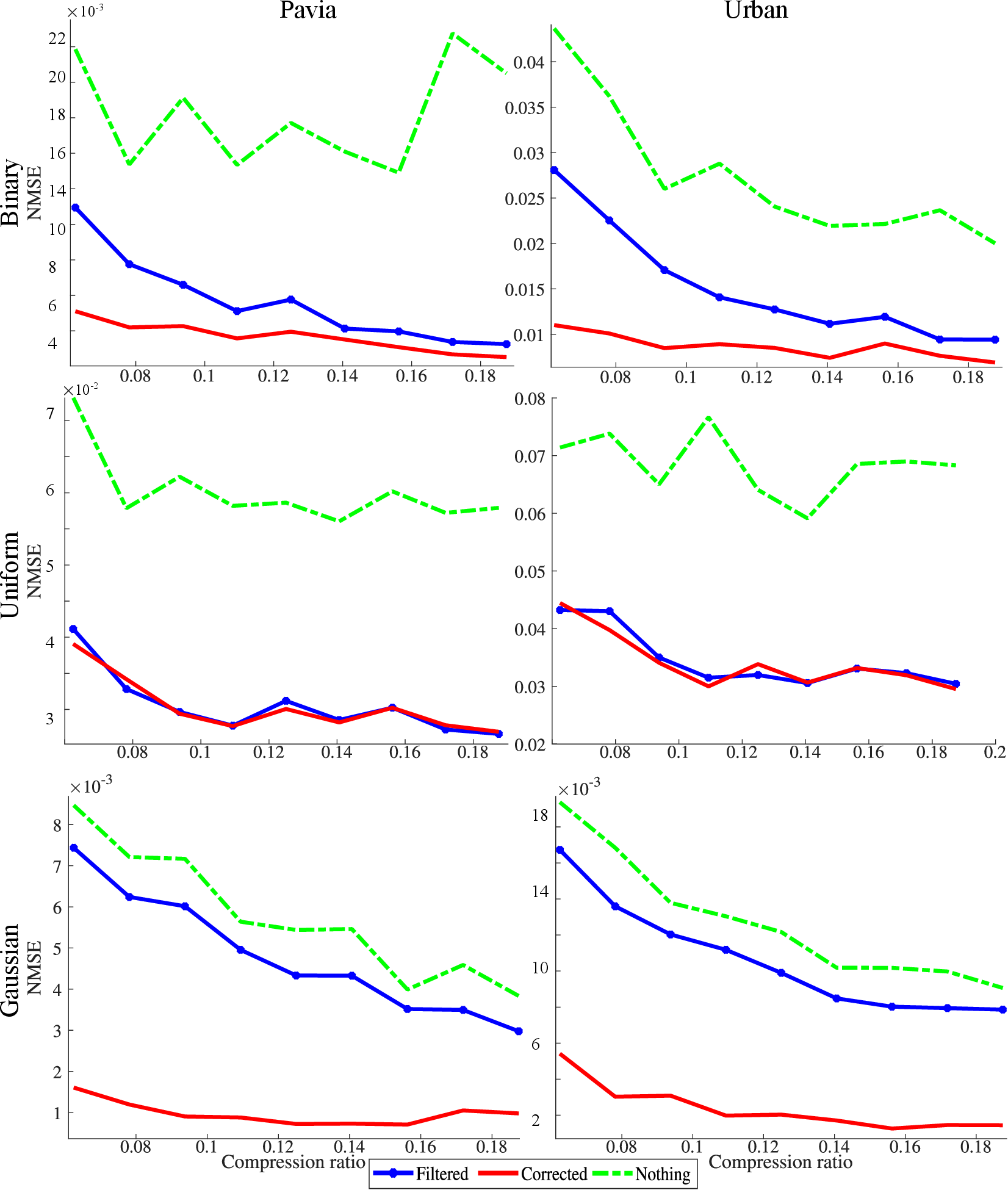}
    \caption{NMSE of the recovered covariance matrix when the filtered gradient is used. Dotted green line represents the unfiltered gradient results. Blue solid lines with dot markers show the results with filtered gradient. Red solid line presents the results of filtered gradient when the error term  \eqref{eq:bias} is subtracted in each gradient step.}
    \label{fig:biascov1}
\end{figure}

Additionally, to test the impact of the error term given in lemma \ref{lema:bias}, this term is computed and subtracted from the gradient and the results shown in Fig. \ref{fig:biascov1}. Results show an important improvement when the filtered gradient is used specially using uniform sensing matrices. Additionally, when the error term is computed (only for comparison purposes) the improvement is up to one order of magnitude, note that this is only possible if the covariance matrix is known beforehand; in the case of the Uniform matrices the simulations shows no improvement when the error term is subtracted. However, the error always decrease using the filtered gradient and the improvement is larger when using Uniform matrices. Additionally, the stability of the reconstruction appears to improve as well. 

\section{Noise analysis}\label{sec:noise}
{\color{black}
	     Several simulations varying the noise level were performed to test how robust to noise the proposed algorithm is. The Figure \ref{fig:noise} shows the NMSE of the reconstructed covariance matrix varying the SNR from 40 dB to 15 dB. It can be seen that the proposed algorithm outperforms all methods when high levels of noise are present in the measurements. For instance, in the Gaussian scenario, the NMSE does not vary that much even though the noise level increases. SpeCA algorithm obtains better results using Gaussian matrices but only for low levels of noise.
	     
	     \begin{figure}[!htb]
	         \centering
	         \includegraphics[width=0.38\textwidth]{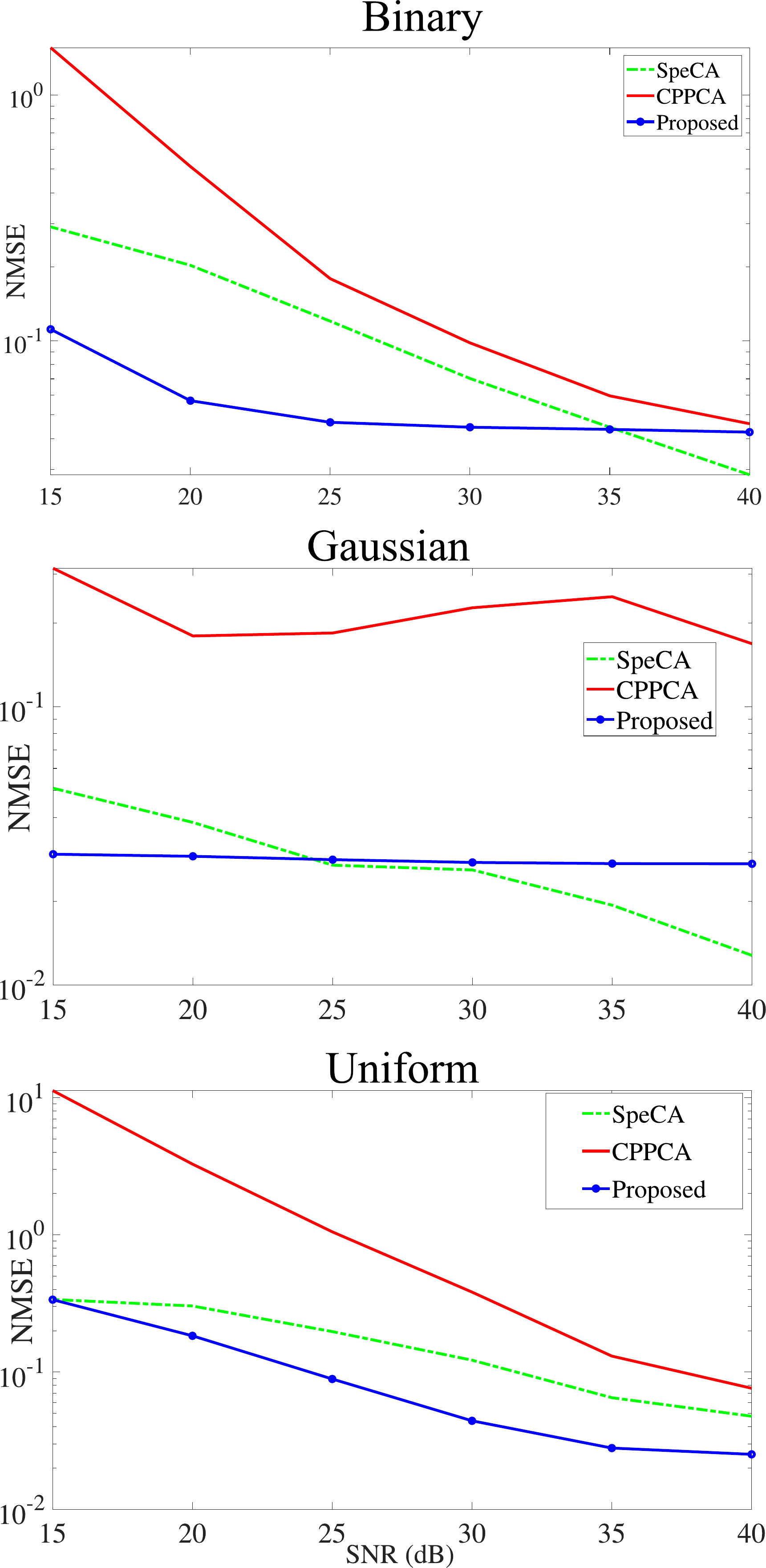}
	         \caption{NMSE of the reconstructed covariance matrix by varying the noise levels for 8\% of compression ratio. Top: Results for Binary sensing matrices. Middle: Results for Gaussian sensing matrices. Bottom: Results for Gaussian Uniform matrices }
	         \label{fig:noise}
	     \end{figure}
	    }
\section{Choosing $\tau$ parameter.}\label{sec:tau}
{\color{black}
	        The optimal parameter should be chosen as $\tau=\rho \text{trace}(\bm{\Sigma})$, where $\rho \in [0,1]$ and depends on the rank of the covariance matrix and variance of the noise. Since $\bm{\Sigma}$ is unknown we used the initialization $\mathbf{S}_0$ of the algorithm in order to approximate it as
	        \begin{equation}
	            \mathbf{S}_0 = \frac{1}{p} \sum_{i=1}^p (\mathbf{P}_i^T)^\dagger \mathbf{\tilde{S}} (\mathbf{P}_i)^\dagger,
	        \end{equation}
	        where $\dagger$ represents the Moore-Penrose pseudo-inverse. The figure \ref{fig:rho} shows the NMSE of the reconstructed covariance matrix by varying the $\rho$ parameter for two levels of noise (30 dB, 20 dB). It can be seen that the  
	        \begin{figure}[!htb]
	            \centering
	            \includegraphics[width=0.40\textwidth]{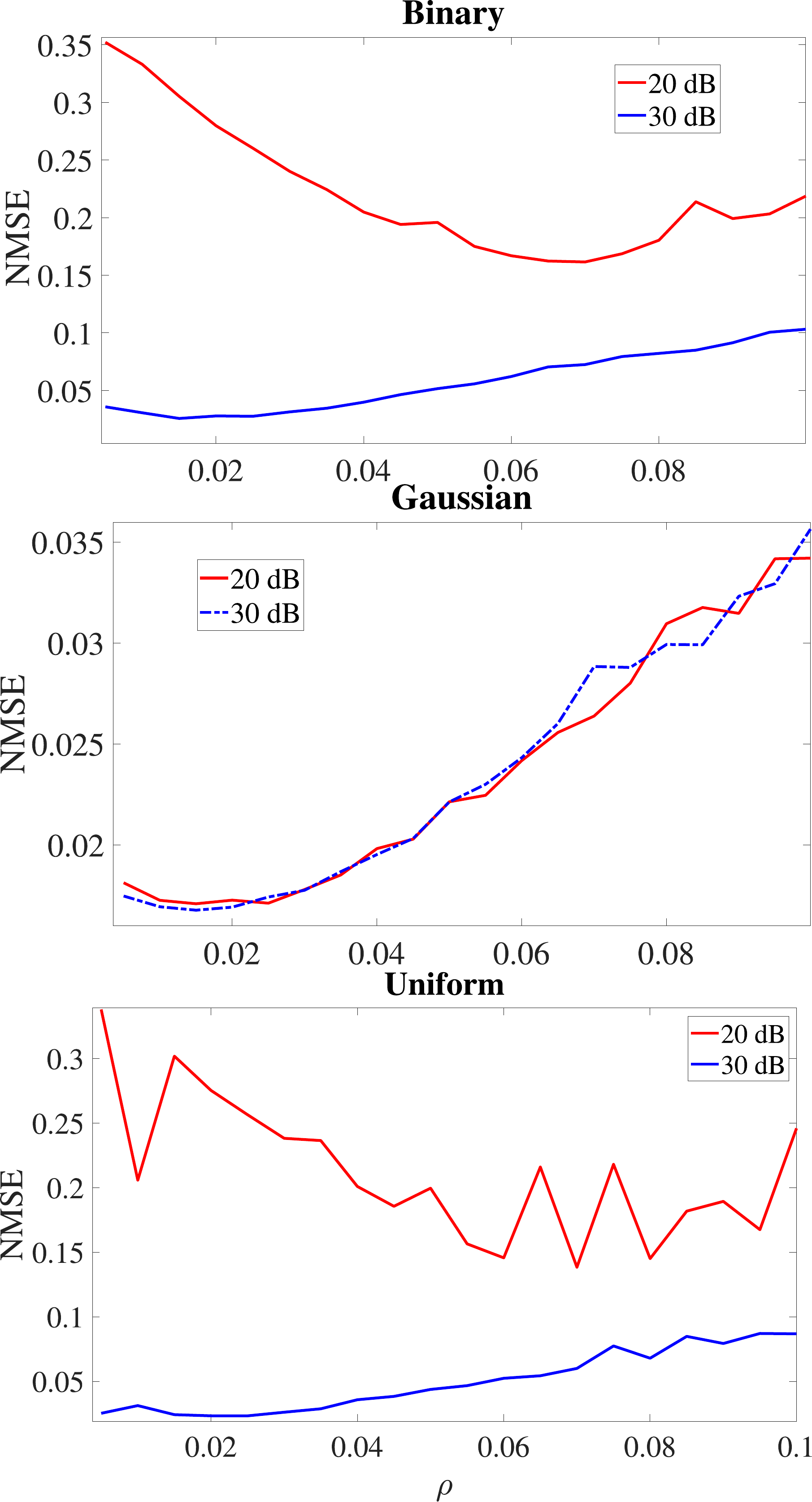}
	            \caption{NMSE of the reconstruction by varying the $\rho$ parameter. Top: NMSE for Binary matrices. Middle: NMSE for Gaussian matrices. Bottom: NMSE for Uniform matrices}
	            \label{fig:rho}
	        \end{figure}
	    }
\section{Convergence analysis}\label{sec:convergence}
This section studies the convergence properties of the projected gradient Algorithm \ref{alg:PG} to solve \eqref{eq:opt3}. Consider the function $\psi(\bm{\Sigma})= \text{Tr}(\bm{\Sigma})$ in \eqref{eq:opt3}, and let
\begin{equation}
        g(\bm{\Sigma}) = \sum_{i=1}^p||\text{vec}(\mathbf{\tilde{S}}_i) -\mathbf{Q}_i\text{vec}(\bm{\Sigma})||_2^2 + \tau \mathbf{d}^T\text{vec}(\bm{\Sigma}) ,
        \label{eq:kron}
\end{equation}
be the vector formulation of \eqref{eq:f}, with $\mathbf{Q}_i=\mathbf{P}_i^T \otimes \mathbf{P}_i^T$, where $\otimes$ is the Kronecker product, vec($\cdot$) denotes the operation that stacks the columns of a given matrix into a column vector, $\mathbf{d} = \text{vec}(\mathbf{I})$ with $\mathbf{I}\in \mathbb{R}^{l\times l}$ the $l\times l$ identity matrix, and $||\cdot||_2$ is the $\ell_2$ norm.  Given that $g(\bm{\Sigma}) \equiv f(\bm{\Sigma})$\cite{Bioucas2014Covalsa}, the function $g(\bm{\Sigma})$ is considered instead of $f(\bm{\Sigma})$,   
\begin{equation}
	\begin{aligned}
	\bm{\Sigma}^* = &\underset{\bm{\Sigma}\in \mathbb{R}^{l\times l}}{\text{ argmin}}
	& & g(\bm{\Sigma}) + h(\bm{\Sigma}),
	\end{aligned}
	\label{eq:opt5}
\end{equation}
where $h(\bm{\Sigma})$ is an indicator function of the positive semi-definitive and Toeplitz set. For simplicity, and taking into account that the function $g(\bm{\Sigma})$ vectorizes the input matrix, both $g(\bm{\Sigma})$ and $g(\bm{\sigma})$ will be used indistinctly, where $\bm{\sigma} = \text{vec}(\bm{\Sigma})$ and $\mathbf{\tilde{s}} = \text{vec}(\mathbf{\tilde{S}})$. Further, Fej\'er proved that sequence of points generated by the projected gradient converges to a solution\cite[Theorem 10.23]{Beck2017}.
\setcounter{section}{10}
\setcounter{theorem}{22}
\begin{theorem}[Fej\'er monoticity theorem]
    Suppose that $g(\bm{\Sigma})$ and $h(\bm{\Sigma})$ are proper closed and convex functions, additionally, dom($h(\bm{\Sigma})$)  $\subseteq$  int(dom($g(\bm{\Sigma})$)) and $g(\bm{\Sigma})$ is L-\textit{smooth}. Let $\{\bm{\Sigma}^k \}_{k \geq 0}$ be the sequence of points generated by the projected gradient algorithm. Then for any optimal point $\bm{\Sigma}^*$ and $k\geq 0$ it holds that
    \begin{equation}
        ||\bm{\Sigma}^{k+1} - \bm{\Sigma}^*|| \leq ||\bm{\Sigma}^{k} - \bm{\Sigma}^*||.
    \end{equation}
    \label{the:convergence}
\end{theorem}
\vspace*{-0.5cm}
\setcounter{section}{4}
\setcounter{theorem}{0}
It can be seen that \eqref{eq:opt5} is convex and differentiable, thus the first assumption of theorem \ref{the:convergence} is accomplished. Additionally, a function $g$ is said to be L-smooth if it is differentiable and there exists $L>0$ such that $||\nabla g(\mathbf{x}) - \nabla g(\mathbf{y}) ||_2 \leq L ||\mathbf{x}-\mathbf{y}||_2$\cite{Beck2017},
for all $\mathbf{x},\mathbf{y} \in E$, with $E$ the domain of the function $g$. Thus, for the function $g(\bm{\Sigma})$ defined in \eqref{eq:kron}, the gradient is given by $
    \nabla g(\mathbf{x}) = \sum_{i=1}^p\mathbf{Q}_i^T(\mathbf{Q}_i\mathbf{x}-\mathbf{\tilde{s}}) + \tau \mathbf{d}.
    $ Plugging it in $||\nabla g(\mathbf{x}) - \nabla g(\mathbf{y}) ||_2$ yields
\begin{equation}
    \left|\left|\left(\sum_{i=1}^p\mathbf{Q}_i^T(\mathbf{Q}_i\mathbf{x}-\mathbf{\tilde{s}}_i) + \tau \mathbf{d}\right) -\left( \sum_{i=1}^p\mathbf{Q}_i^T(\mathbf{Q}_i\mathbf{y}-\mathbf{\tilde{s}}_i) + \tau \mathbf{d}\right) \right|\right|_2,
    \label{eq:24}
\end{equation}
and after some algebraic manipulations, \eqref{eq:24} can be rewritten as
\begin{equation}
     \left|\left|\sum_{i=1}^p(\mathbf{Q}_i^T\mathbf{Q}_i)(\mathbf{x}-\mathbf{y})\right|\right|_2 \leq \left|\left|\sum_{i=1}^p(\mathbf{Q}_i^T\mathbf{Q}_i)\right|\right|\ ||\mathbf{x}-\mathbf{y}||_2.
\end{equation}
This implies that $L=||\sum_{i=1}^p(\mathbf{Q}_i^T\mathbf{Q}_i)||$. Thus, it can be concluded that the sequence of points generated by Algorithm \ref{alg:PG} is Fej\'er monotone which guarantees convergence.
{\color{black}
\section{Incidence of the kernel size of the filtering gradient in the estimation accuracy}
The kernel size in the filtering step affects the reconstruction accuracy since, as shown in Fig. \ref{fig:biascov}, no filtering of the gradient results in poor reconstruction in the eigenvectors associated with the smallest recovered eigenvalues. On the other hand, an over-smoothed gradient can also affect the reconstruction process. To select the correct kernel size $k$ and variance $\sigma$, we perform multiple simulations to show the error versus the variance of the Gaussian kernel. The kernel size was chosen following $k=2 \lceil 2\sigma \rceil +1$. From Fig. \ref{fig:kernel}, it can be seen that the optimal variance value is $\sigma=1$. 
\begin{figure}[!htb]
    \centering
    \includegraphics[width=\linewidth]{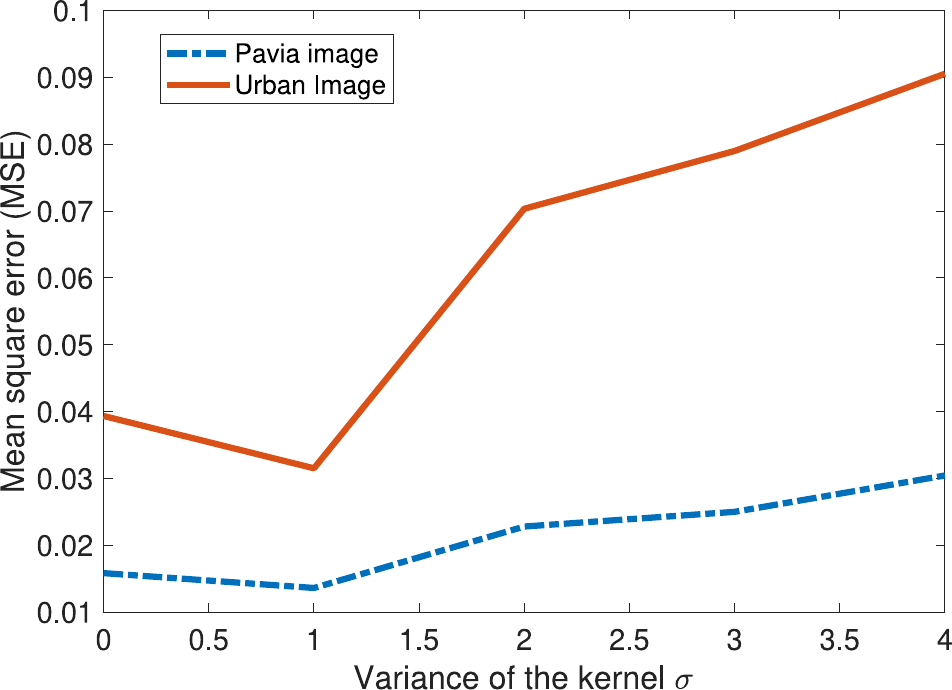}
    \caption{Reconstruction error by varying the variance and size of the Gaussian kernel. Note that the kernel size was set to $2 \lceil 2\sigma \rceil +1$.}
    \label{fig:kernel}
\end{figure}
}
\bibliographystyle{ieeetr}
\bibliography{main}